\documentclass{article}

\usepackage{arxiv}

\usepackage{amssymb}
\usepackage[utf8]{inputenc}
\usepackage{natbib}
\usepackage{subcaption}
\usepackage{booktabs}
\usepackage{hyperref}
\usepackage{placeins}
\usepackage[capitalise,poorman]{cleveref}
\usepackage[pdftex,dvipsnames]{xcolor}  %
\usepackage[colorinlistoftodos,prependcaption,textsize=tiny]{todonotes}
\usepackage{tikz}
\usetikzlibrary{backgrounds,shadows,scopes,fit,calc,patterns}

\newcommand{\extendnoderight}[1]{
    (#1)
    ($(current bounding box.north east)!(#1)!(current bounding box.south east)$)
}
\newcommand{\extendnodeleft}[1]{
    (#1)
    ($(current bounding box.north west)!(#1)!(current bounding box.south west)$)
}

\title{Perceptions of Edinburgh: Capturing Neighbourhood Characteristics by Clustering Geoparsed Local News}

\usepackage{authblk}

\author[1]{%
  Andreas Grivas\thanks{\texttt{agrivas@ed.ac.uk}}%
}
\author[1]{%
  Claire Grover%
}
\author[1]{Richard Tobin}
\author[2,5]{Clare Llewellyn}
\author[2]{Eleojo Oluwaseun Abubakar}
\author[3]{Chunyu Zheng}
\author[3]{Chris Dibben}
\author[2]{Alan Marshall}
\author[3]{Jamie Pearce}
\author[4,5]{Beatrice Alex}

\affil[1]{School of Informatics, University of Edinburgh, Edinburgh, EH8 9AB}
\affil[2]{School of Social and Political Science, University of Edinburgh, Edinburgh, EH8 9LD}
\affil[3]{School of Geosciences, University of Edinburgh, Edinburgh, EH8 9XP}
\affil[4]{School of Literatures, Languages and Cultures, University of Edinburgh, Edinburgh, EH8 9JU}
\affil[5]{Edinburgh Futures Institute, University of Edinburgh, Edinburgh, EH3 9EF}

\begin{document}

\maketitle

\begin{abstract}
The communities that we live in affect our health in ways that are complex and hard to define.
Moreover, our understanding of the place-based processes affecting health and inequalities is limited.
This undermines the development of robust policy interventions to improve local health and well-being.
News media provides social and community information that may be useful in health studies.
Here we propose a methodology for characterising neighbourhoods by using local news articles.
More specifically, we show how we can use Natural Language Processing (NLP) to unlock further information about neighbourhoods by analysing, geoparsing and clustering news articles.
Our work is novel because we combine street-level geoparsing tailored to the locality with clustering of full news articles, enabling a more detailed examination of neighbourhood characteristics.
We evaluate our outputs and show via a confluence of evidence, both from a qualitative and a quantitative perspective, that the themes we extract from news articles are sensible and reflect many characteristics of the real world.
This is significant because it allows us to better understand the effects of neighbourhoods on health.
Our findings on neighbourhood characterisation using news data
will support a new generation of place-based research which examines a wider set of spatial processes and how they affect health, enabling new epidemiological research.
\end{abstract}

\section{Introduction}
The communities that we live in affect our health in ways that are complex and hard to define \citep{baciu2017root}. Neighbourhoods influence population health, and while some characteristics are captured by indexes, such as the Scottish Index of Multiple Deprivation (SIMD)\footnote{\url{https://simd.scot}}, they are expensive to produce, and can quickly become out of date. Moreover, these statistics cannot always tell the full story, since factors like social support, antisocial behaviour and community cohesion are challenging to define, measure, and capture. Such omissions limit our understanding of the place-based processes that affect health and inequalities and undermine the development of robust policy interventions to improve local health and wellbeing.

However, a lot of neighbourhood information and knowledge on communities is captured in an ephemeral manner in local news and on social media and this sea of textual information remains relatively untapped.
We hypothesise that local news captures social and community information that may be useful in health studies.
As such, we propose a simple methodology for characterising neighbourhoods by leveraging local newspaper articles, in our case, articles which report on areas in the City of Edinburgh and span four years between 2017-2021.
We demonstrate how Natural Language Processing (NLP) algorithms which rapidly analyse, geoparse and cluster news articles can be used to extract such neighbourhood characteristics.
The insights gained from these characteristics can be used by social scientists and geographers to further examine and understand neighbourhood health.

An observation we would like to better understand using our work is the presence of neighbourhoods that are more (or less) resilient to ill health than would be expected.
Such expectations are formed from information recorded in existing statistical information on mortality and deprivation \citep{pearson2013deprived}, and via qualitative research using in-depth interviews and surveys \citep{mitchell2009factors, popay2022system}.
Our research aims to support research on identifying which characteristics of neighbourhoods increase vulnerability or promote resilience to ill health.

The outline of this paper is as follows.
We first discuss related work in this area (\cref{section:related_work}).
We then describe the creation of the dataset used for this research (\cref{section:data}).
In~\cref{section:methodology}, we provide a detailed overview of our methodology for characterising neighbourhoods by combining geoparsing with hierarchical clustering of news articles.
In~\cref{section:evaluation} we describe our qualitative and quantitative clustering evaluation. We conclude with a discussion of our results in~\cref{sec:discussion} and future work in~\cref{sec:future}.

\section{Related Work}\label{section:related_work}

Our work adds to innovative research which apply NLP techniques, such as geoparsing and clustering, on news data to better understand neighbourhood characteristics.
This is an important line of research since the relationship between neighbourhood characteristics and health outcomes is understudied.
Our work takes inspiration and guidance from the related ideas and techniques presented in multiple domains.

\subsection{Capturing Location Characteristics}
\citet{Lansley2016} analysed how the content Twitter users post online varies across neighbourhoods and demographics.
To do so, they applied topic modelling to geo-referenced social media data in the form of geo-tagged Tweets from Inner London. The motivation of their work was to use the classification of ephemeral social media as an alternative source of information to public surveys, to characterise place and demonstrate how the nature of such content varies by place and user. They used Latent Dirichlet Allocation (LDA) \citep{blei2003latent} to extract topics from a large number of Tweets. LDA requires the manual setting of topic number. We apply a hierarchical clustering to model topics in news data which has the advantage that it automatically derives the number of clusters and sub-clusters (see Section~\ref{section:methodology}). \citet{Lansley2016} manually labelled their 20 derived topics and presented their distributions over time of day, per location in Inner London (residential versus non-domestic buildings or public green spaces as well as per six key locations) and by gender of Tweet authors. We use a similar approach in using clustered data in our methodology.
However, we also evaluate the quality of our clusters by annotating pairs of articles and use the annotations to select the clustering parameters in our NLP pipeline.
Our NLP-derived neighbourhood profiles will be used as input in the follow-on analysis of neighbourhood health outcomes.

~\cite{bechini2022news} describe a news-based framework for uncovering and tracking city area profiles that are used to show how COVID-19 affected citizens in Rome. They derived area profiles by applying the clustering of titles and summaries of news articles in combination with geo-referencing location metadata \citep{bondielli2020exploiting}.  They proposed a clustering method that accounts for changes over time. They used the clustering of news data related to the city of Rome as a case study, examined their results qualitatively. They also introduced metrics to analyse how clusters changed from one time window to the next. Their geo-referencing uses location mentions provided in the meta-data of the collected news. The areas within Rome are divided into 1km-by-1km squares, as opposed to defining neighbourhoods as understood by those living within the city. In contrast, our analysis required the examination of news relevant to existing neighbourhoods within Edinburgh as defined by humans, and the application of geo-referencing and clustering to the entirety of the news article. In this paper, we took our quantitative evaluation of clustering a step further. We did so both by evaluating how well our clustering grouped related articles together, as well as by comparing our results to known statistics.

\citet{zolnoori2021mining} combined topic modelling and sentiment analysis of news data to determine public opinion communicated towards health concerns. They identified a list of 10 major public health concerns by analysing MeSH terms, a controlled terminology created by the National Library of Medicine, from 30 leading public health journal articles \citep{lipscomb2000medical}. They then collected news articles published by Reuters between 2007 and 2017 using a news crawler and filtered them for the top 10 public health concerns.  They applied topic modelling using Topic Keyword Model (TKM) \citep{schneider2018topic} and sentiment analysis using a lexicon-based approach (VADER) \citep{hutto2014vader}. They compared their topic trends over time and plotted against Google trends searches rather than conducting a formal evaluation.

\subsection{Information Extraction from News Data}
There is also a large body of research on applying information extraction extract insights from news data. Highlights of this include \citet{dasgupta2017crimeprofiler}, who use computational linguistics-based methods for detecting information related to crime in news.  Geographical Text Analysis refers to a combination of corpus linguistics, critical discourse analysis and geographical information systems conducted on historical corpora and contemporary newspapers in the context of place, poverty and health \citep{porter2015geographical,porter2018space,oro58565,paterson2018representations}. \cite{porter2018space} analysed articles published in a London-based newspaper in the nineteenth century for mentions of disease. \cite{paterson2018representations} analysed contemporary news articles from the Guardian and the Daily Mail published in 2010-2015. They applied a keyword search for the term poverty to identify and geo-reference snippets of news text. They then applied density smoothing to map the number of mentions of poverty for places located in the United Kingdom. \cite{oro58565} conducted similar work related to poverty by analysing articles published between 1940 and 2010 in The Times.  These studies conducted targeted keyword-based analysis of specific terms related to health, mortality, and poverty to plot the frequency of mentions and map them to geographical locations mentioned in context.  These findings inspired our examination of news data for topics related to community cohesion which affect neighbourhood health outcomes and are not captured effectively in statistical information collected in health or deprivation indexes like SIMD.

\section{News Data}\label{section:data}

\subsection{Collection}
We focussed on news data that was relevant to, and representative of, local communities within the City of Edinburgh, UK. The aim was to understand the neighbourhood and the activities that occurred within that micro-level of community. We decided that local news would best represent content at this level.  We therefore gathered articles from the Edinburgh Evening News\footnote{\href{https://www.edinburghnews.scotsman.com/}{https://www.edinburghnews.scotsman.com/}}
published between 2017 and 2020 using the Lexis Nexis platform API\footnote{\href{https://www.lexisnexis.co.uk/}{https://www.lexisnexis.co.uk/}} via the University of Edinburgh Library. This resulted in 72,700 retrieved articles in PDF format (see Table~\ref{tab:datastats}). As can be seen, there were relatively few articles available from the second half of 2019, and almost none from the first half of 2020, the reason for this is not given and it is unclear why this is the case.

\subsection{Deduplication}
Articles in digital news are often revised and republished. As such, our retrieved articles included many duplicates and near-duplicates. We considered articles to be duplicates if they had at least half their sentences in common. We ignored sentences with fewer than 10 words and sentences that occurred in more than 20 articles, as many of these were ``boilerplate'' and did not imply duplication. Among the duplicates we selected the longest article. This left 66,601 unique articles (see Table~\ref{tab:datastats} for a breakdown per year).

\begin{table}
\begin{center}
\begin{tabular}{|c|c|c|}
\hline
Year & Articles retrieved & Unique articles \\
\hline
2017 & 15,710 & 13,962 \\
2018 & 17,862 & 15,574 \\
2019 & 11,347 & 9,593 \\
2020 & 7,603 & 7,478 \\
2021 & 20,178 & 19,994 \\
\hline
Total & 72,700 & 66,601 \\
\hline
\end{tabular}
\end{center}
\caption{Data statistics}
\label{tab:datastats}
\end{table}

\section{Natural Language Processing Methodology}\label{section:methodology}
Our overall goal in this work is to use news data to learn more about neighbourhoods.
In our approach, we will characterise neighbourhoods based on the topics that are present in the articles that mention them.
To this end, in the following steps, we:
\begin{itemize}
\item Identify location mentions in each article.
\item Cluster the news articles into topics and themes.
\item Characterise each location by computing its distribution of topics.
\end{itemize}

\begin{figure}[t]
\centering
\tikzset{%
    color node/.style={%
        minimum height=1.5cm, minimum width=2cm, draw, rectangle, inner color=#1!10, outer color=#1!30, align=center, outer sep=.08cm, font=\small
    },
    title node/.style={%
    	rectangle, font=\large\itshape, anchor=west, align=center, color=#1
    },
    label node/.style={%
    	rectangle, font=\small\itshape, anchor=west, align=center
    },
    text node/.style={%
    	rectangle, font=\Large\ttfamily, anchor=west, align=center, color=#1
    },
    myarrow/.style={%
    	->, thick, draw=Black
    },
    myarrowlbl/.style={%
    	font=\itshape, midway, sloped, align=center, above
    }
}

\begin{tikzpicture}
    \def\x{0}%
    \def\y{0}
    \def\bs{.1}
    \def\offset{.8}
    \node [color node={RoyalBlue}] (geotag) at (\x, \y) {Identify\\ Location};
    \node [color node={RoyalBlue}, below=.5*\offset of geotag] (cluster) {Cluster};
    \node [color node={RoyalBlue}, right=\offset of cluster, yshift=1cm] (profile) {Characterise};
    
    \begin{scope}[on background layer]
     \node (backweb) [fill=RoyalBlue!15,rounded corners,fit=\extendnoderight{profile}\extendnodeleft{geotag}\extendnodeleft{cluster}] {};
    \end{scope}
    
    \node [left=1.5*\offset of geotag, yshift=-1cm] (input) {\includegraphics[width=.2\textwidth]{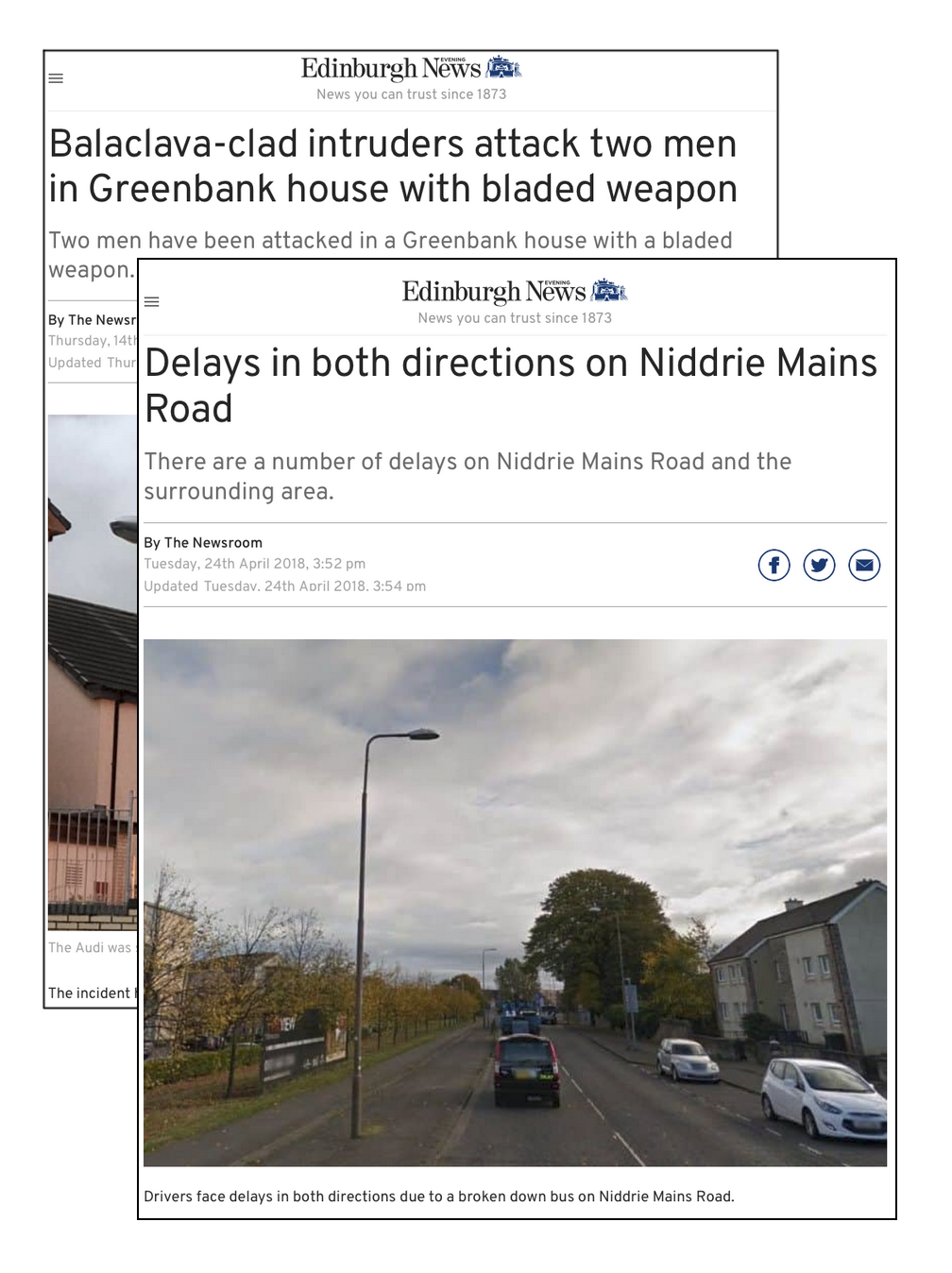}};
    \node [right=\offset of profile] (output) {\includegraphics[width=.18\textwidth]{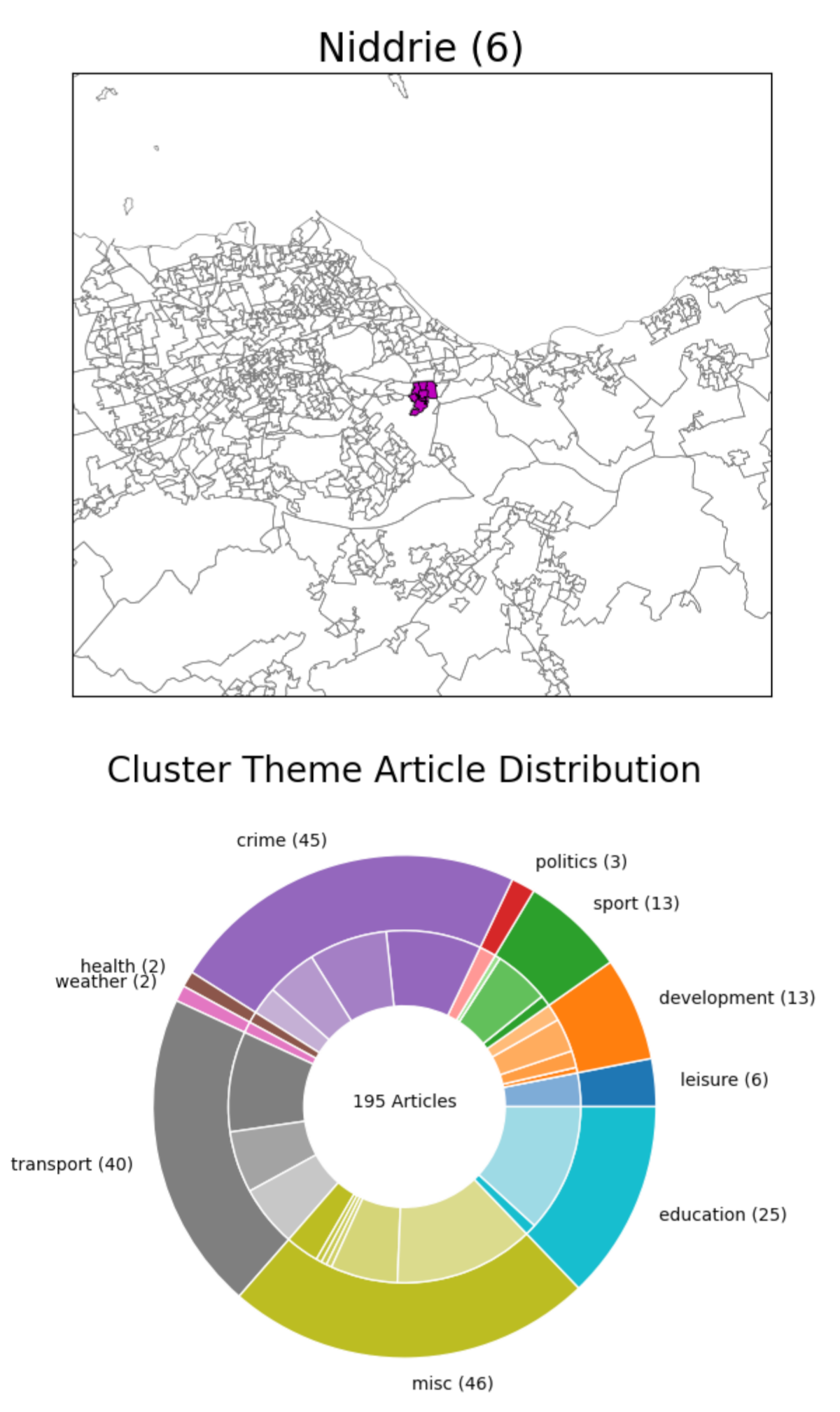}};
    
    \draw[myarrow] (input.east) to (geotag.west);
    \draw[myarrow] (input.east) to (cluster.west);
    \draw[myarrow] (profile.east) to (output.west);
    \draw[myarrow] (geotag.east) to[out=10, in=90] (profile.north);
    \draw[myarrow] (cluster.east) to[out=350, in=270] (profile.south);
    
    \node [title node={Black}, above=\offset*.2 of input] (ilbl) {News Articles};
    \node [title node={RoyalBlue}, above=\offset*.28 of geotag, xshift=1.5cm] (lbl) {News Clustering};
    \node [title node={Black}, above=-\offset*.2 of output] (olbl) {Summary};
    
\end{tikzpicture}
\caption{Given a dataset of Edinburgh news articles (N=66,601) we a) identify which locations are mentioned in the articles and b) cluster the articles into themes. We then aggregate the cluster information by location and summarise neighbourhoods as a distribution over themes.}
\label{fig:pipeline}
\end{figure}
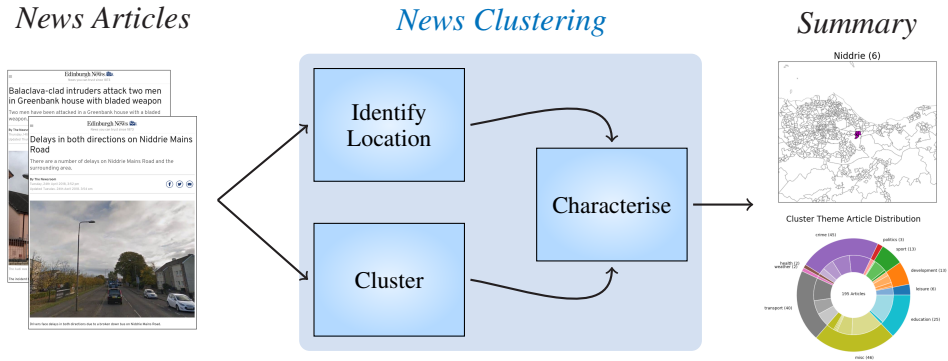

We illustrate our approach in~\cref{fig:pipeline}.
There are several key concepts represented in this work that are worth exploring and defining in more detail:
\paragraph{Location}
In this paper, we will focus our analysis on Edinburgh and its surrounding areas, since these are the locations covered in our news articles.
We will discuss locations at two granularities.
The most fine grained locations we have statistics for are data zones, as used in the 2020 Scottish Index of Multiple Deprivation (SIMD).\footnote{\href{https://www.gov.scot/collections/scottish-index-of-multiple-deprivation-2020/}{https://www.gov.scot/collections/scottish-index-of-multiple-deprivation-2020/}}
A \textbf{data zone} is a small area that has roughly 500 to 1000 household residents.
They are large enough such that statistics can be represented accurately without loss of privacy, and yet small enough to represent distinct geographical communities.
Scotland is divided into 6976 data zones based on the 2011 census and defined by polygons, with rural ones being larger in area.
We will evaluate our clusters on the data zone granularity, since that is the granularity at which statistics in the SIMD are reported.
However, for reasons we will discuss in~\cref{sec:discussion}, we will use our clustering outputs to characterise coarser areas than data zones.
We will call such a coarser area a \textbf{neighbourhood}.
A neighbourhood is the union of all data zones having the same name.

\paragraph{Article Groups} A \textbf{cluster} is a group of news articles that are related, in the sense that they talk about similar things. In general, clustering methods do not label the clusters they extract, so it is left to us to interpret them. A \textbf{topic} is a cluster which we have named, such as ``football'', ``tennis'' or ``thefts''. A \textbf{theme} is a group of topics, such as ``sports'' or ``crime''.

\subsection{Location and Area Identification}

In order to extract locations as defined by a latitude and longitude from text we use a geoparser. For this work, to deal with finer granularity of location, we used a specially customised version of the \href{https://www.ltg.ed.ac.uk/software/geoparser/}{Edinburgh Geoparser}~\citep{Alex2017} to identify Edinburgh locations in the news articles. This comprises the following sequence of processing steps:

\paragraph{Preprocessing} The news articles are converted from PDF to XML and are further manipulated to find the body of the text and paragraph breaks within it as well as the title of the article. Some parts of the PDF, such as adverts, are discarded while other parts are placed in metadata fields including date of publication, bylines, highlights, section tags, and links. The text body is processed to identify sentences and tokens and to part-of-speech tag and lemmatise words.

\paragraph{Named Entity Recognition} Unlike the standard Geoparser, we aimed for a finer granularity of location named entities: as well as town, village and area names we identified smaller locations such as names of streets, buildings, parks, churches and schools. To this end we refined parts of the rule-based Named Entity Recognition component by adapting rules and adding an Edinburgh-specific lexicon of place names derived from  \href{https://www.ordnancesurvey.co.uk/products/os-open-names}{Ordnance Survey Open Names}, the most comprehensive database of names.

\paragraph{Georesolution} The standard Geoparser assigns latitude/longitude coordinates to all location named entities by looking them up in a gazetteer such as \href{https://www.geonames.org/GeoNames}{GeoNames} and resolving any ambiguities arising from multiple possibilities. For this project, we were only interested in places within Edinburgh but we needed a fine-grained local gazetteer. OS Open Names contains coordinates and postcode information so we derived a gazetteer resource by extracting all the entries with an Edinburgh (EH) postcode. This was refined in various ways, e.g.\ by removing duplicates within the same postcode. The georesolution algorithm chooses the most likely Edinburgh gazetteer entry for each place name in the context of the entire article, e.g.\ interpreting ``High Street'' differently in articles about Penicuik and Musselburgh.

\paragraph{Identifying Data Zone Mentions} Georesolved location mentions are mapped to the data zones defined in the \href{https://www.gov.scot/publications/scottish-index-of-multiple-deprivation-2020v2-data-zone-look-up/}{Scottish Index of Multiple Deprivation 2020}. As such, for each article, we obtain a count for the number of times a data zone has been mentioned as well as the corresponding spans of text that prompted this georesolution. These data zone mentions are used to characterise areas according to their distribution of topics from the clustering.

\subsection{News Clustering}
\paragraph{Preprocessing}
Before clustering, we conduct additional pre-processing.
We \emph{lowercase} all text and \emph{remove all space delimited tokens that are longer than 25 characters}.
In order to increase the saliency of the tf-idf vectors, we use SpaCy\footnote{https://spacy.io/}'s part of speech tagging to \emph{remove all determiners, conjunctions, symbols, pronouns, adverbs, adpositions and auxilliary verbs}.
In order to mitigate confounding in clustering (clusters may capture information due to location names), \emph{we mask all location name spans} identified by the geoparser with a placeholder. Moreover, \emph{we ignore broad mentions}, such as that of ``Edinburgh'', since the location name can be used very broadly and its resolution is unlikely to be correct.
Lastly, some articles contain spurious mentions of locations. For example, sports articles or articles announcing railway schedules will name multiple towns, but they do not contain topic information about these locations. As such, \emph{we remove all articles that have more than 40 different location mentions}.

\paragraph{Article Representations}
Since the articles we are clustering are reasonably long, we can rely on word counts to build article representations. We use tf-idf vectors with a vocabulary of $20,000$ and truncated rare tokens, i.e. those for which we had less than $5$ counts. We then use UMAP~\citep{McInnes2018} to reduce the dimensionality of the tf-idf vectors from $20000$ to $10$, using a neighbourhood size of $5$ and the Hellinger metric. We run UMAP for $1000$ epochs. The size of the vocabulary and the UMAP hyperparameters we report here were the ones that worked the best according to our cross-validation, see~\cref{sec:results}.

\paragraph{Clustering}
We use HDBScan~\citep{Campello2013, McInnes2017} with a minimum cluster size of $250$.
HDBScan allows for outliers by including a ``noise'' cluster.
We chose the minimum cluster size such that we get a manageable number of clusters (e.g.\ around 50) while at the same time not assigning too many articles to the noise cluster.

\paragraph{Soft Cluster Assignment}
Since articles may be relevant to more than a single cluster, we rely on the soft scores returned by HDBScan to assign each article to more than one cluster and retain the corresponding affinity scores.
As such, for each article, we obtain $P_{art}(c)$, the probability of an article being assigned the cluster $c$. This is useful for the following step where we want to characterise locations.

\subsection{Characterising Locations}
\paragraph{Cluster Distributions}
We assume that if a location is mentioned in an article, the contents of the article characterise the location in some way.
As a proxy for this characterisation, we use the information returned by the clustering model, namely we characterise a location via a probability distribution over the clusters.
To do so, for each location, $loc$, we keep track of the set of articles, $A_{loc}$, that mentioned this location at least once.
We then aggregate the distributions $P_{art}(c)$ for the articles into a distribution for a location, $P_{loc}(c)$, by summing and normalising.
\begin{equation}
P_{loc}(c) = \frac{1}{|A_{loc}|}\sum_{art \in A_{loc}} P_{art}(c)
\end{equation}

\paragraph{Topics: Naming the Clusters}
A caveat in our characterisations above is that the clustering algorithm will not give us labels for the extracted clusters.
We therefore interpreted the clusters ourselves, see~\ref{app:wclouds}.
We refer to such a named cluster as a \textbf{topic}.
As such, we characterise each location by specifying to what degree different topics are covered in articles that mention this location. For example, a location may be $50\%$ tennis, $20\%$ charity and $30\%$ business.

\paragraph{Grouping the Topics into Themes}
Since many topics end up being related, e.g. both Cluster 2 (Tennis) and Cluster 3 (Football) are sports, we further group topics into themes.
As we will see in~\cref{sec:hier}, this grouping is also present in the cluster hierarchy returned by the model to a large degree, providing further evidence that the returned clusters are semantically meaningful.

\begin{figure}
    \centering
        \begin{subfigure}[b]{.48\textwidth}
        \includegraphics[width=\textwidth]{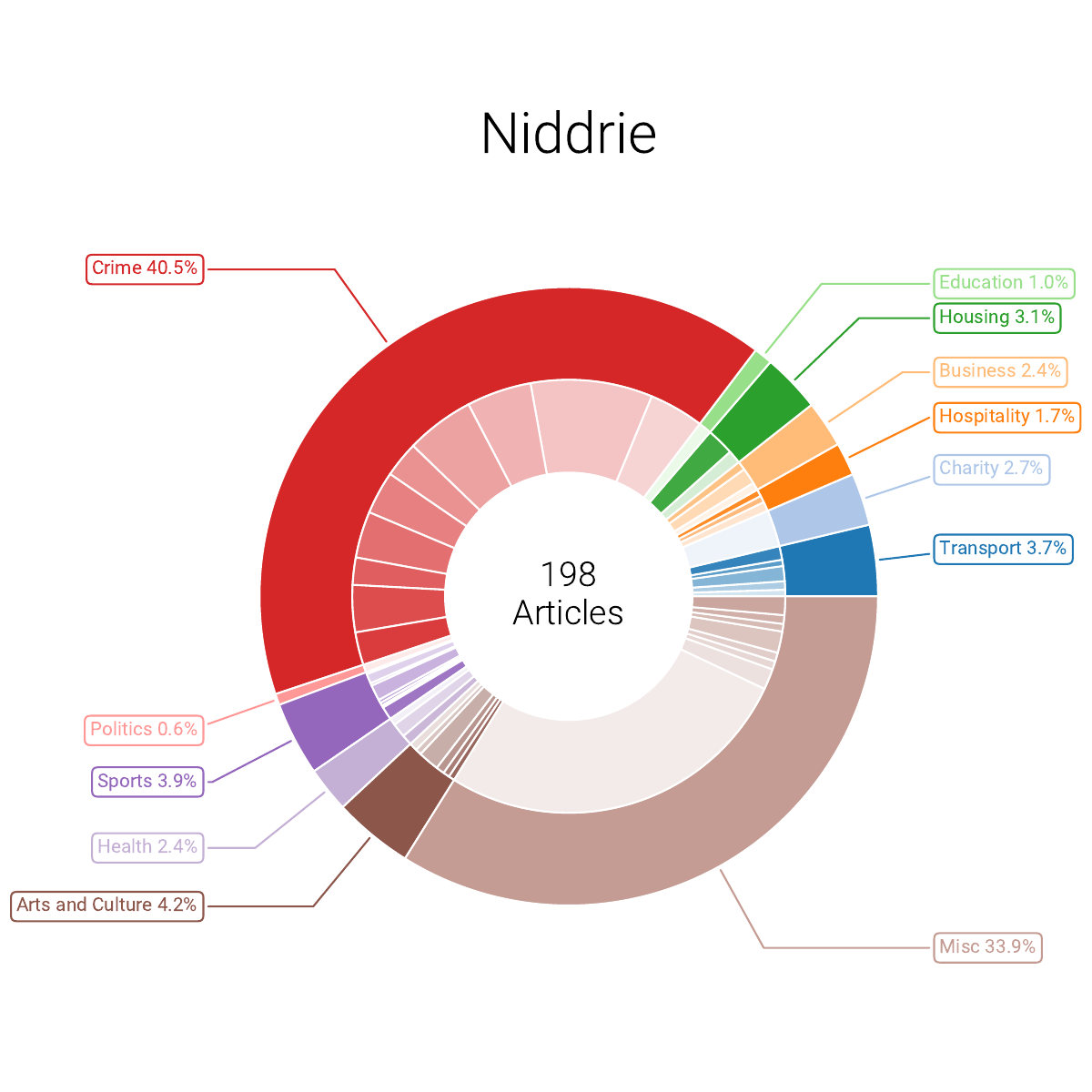}
        \end{subfigure}
        \hfill
        \begin{subfigure}[b]{.48\textwidth}
        \includegraphics[width=\textwidth]{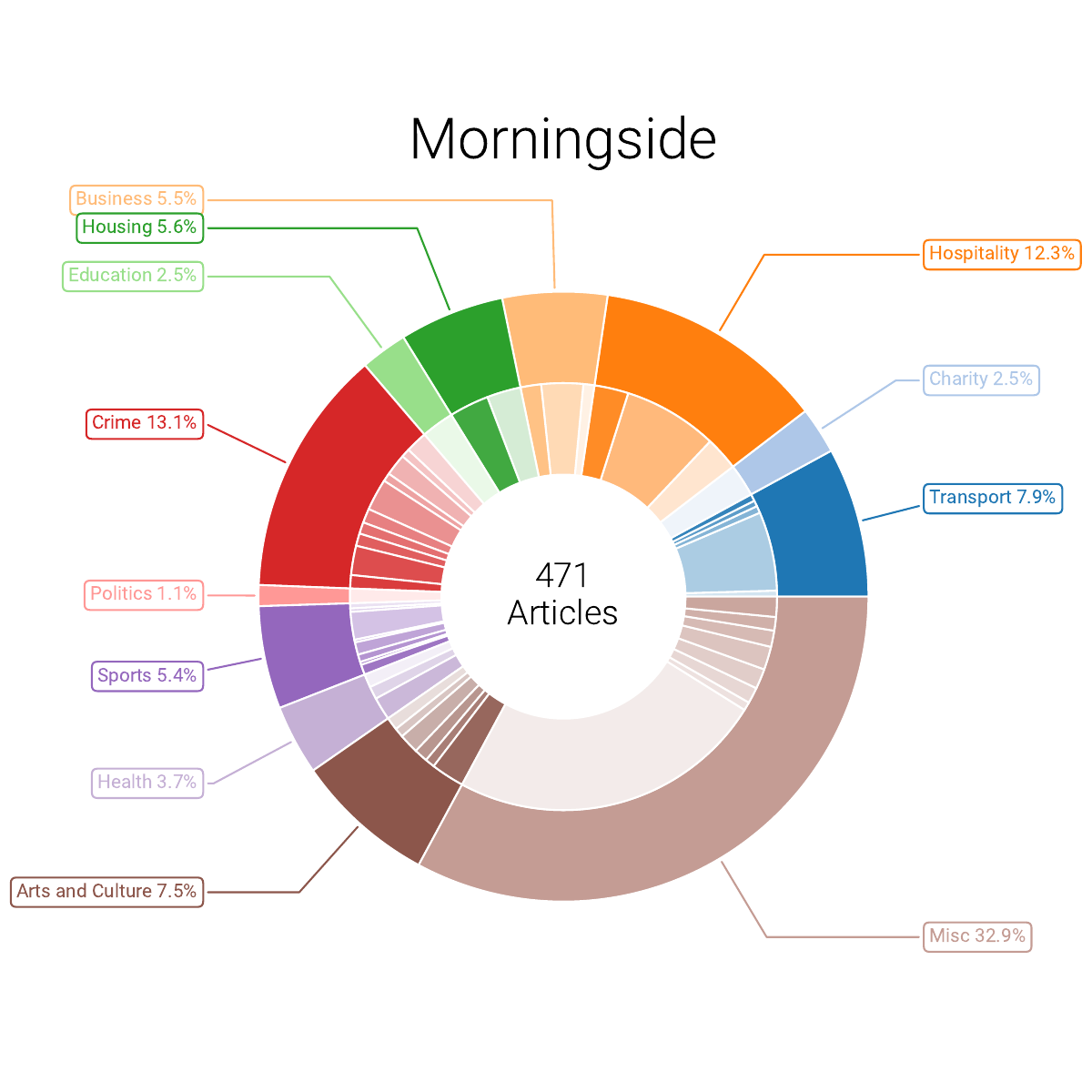}
        \end{subfigure}
    \caption{Characterising two neighbourhoods in terms of their distribution over themes. The distribution over themes is broken down into distributions over topics which can be seen in the inner ring of the chart.}
    \label{fig:pie}
\end{figure}

\paragraph{Neighbourhood Characterisation}
In order to improve the robustness of our characterisations, we conduct our analysis on a more coarse grained level (see~\cref{sec:discussion} for details). To do so, instead of keeping track of mentions of data zones, e.g. ``Oxgangs - 01'' and ``Oxgangs - 02'', we zoom out and keep track of the coarser neighbourhood, such as ``Oxgangs''.
Examples of such characterisations for neighbourhoods can be seen in~\cref{fig:pie}.

\section{Evaluation}\label{section:evaluation}
In this Section we ground our analysis by verifying that our clusters make sense both through a qualitative as well as a quantitative lens.

\subsection{Clustering Annotation}
\label{sec:clusterann}
Clustering evaluation is challenging, because if we knew what the clusters should be we would have used a supervised learning algorithm in the first place.
Herein we take an efficient supervised approach.
We annotate pairs of clusters as to whether they are similar enough to be included in the same cluster or not.
We annotated a subset of $700$ article pairs.
We made sure to include candidate pairs that were likely to end up in the same cluster, by biasing the pairs sampled for annotation towards those that had been assigned the same keywords by the news source.\footnote{Note that if we randomly choose pairs of articles for annotation, it is very unlikely that they will fall in the same cluster.} See~\ref{app:annotation} for more details.
Moreover, we initially categorised the article pairs into 4 strata depending on how similar they were: \textit{Not related}, \textit{Vaguely related}, \textit{Somewhat related} and \textit{Very related}. For simplicity, we binarised the strata into two categories: very related articles should be in the same cluster, while the remaining should not. In~\cref{sec:cluster-eval}, we will use our dataset of annotated article pairs to show that our method indeed creates meaningful clusters.
\subsection{Qualitative Evaluation}

\label{subsec:qual}
We evaluated our clusters qualitatively by inspecting some clusters and the clustering hierarchy.
\paragraph{Inspecting the Clusters}
We inspected the clusters visually and verified that the distributions over topics made sense for some areas of interest.
For example, Cluster 8, which is airport related, has the second largest activation for \textit{Ratho, Ingliston and Gogar - 01}, which is the data zone Edinburgh Airport belongs to.
Moreover, Cluster 4, which is rugby related, has one of its largest activations for \textit{Murrayfield and Ravelston - 01}, where Murrayfield Stadium is situated.
As a last example, Cluster 10, which mentions the trams is relatively active in areas in Leith, which makes sense since the extension of the tram to Leith was approved in $2019$ and began construction in $2020$.

\begin{figure}[h!]
    \centering
    \includegraphics[width=.9\textwidth]{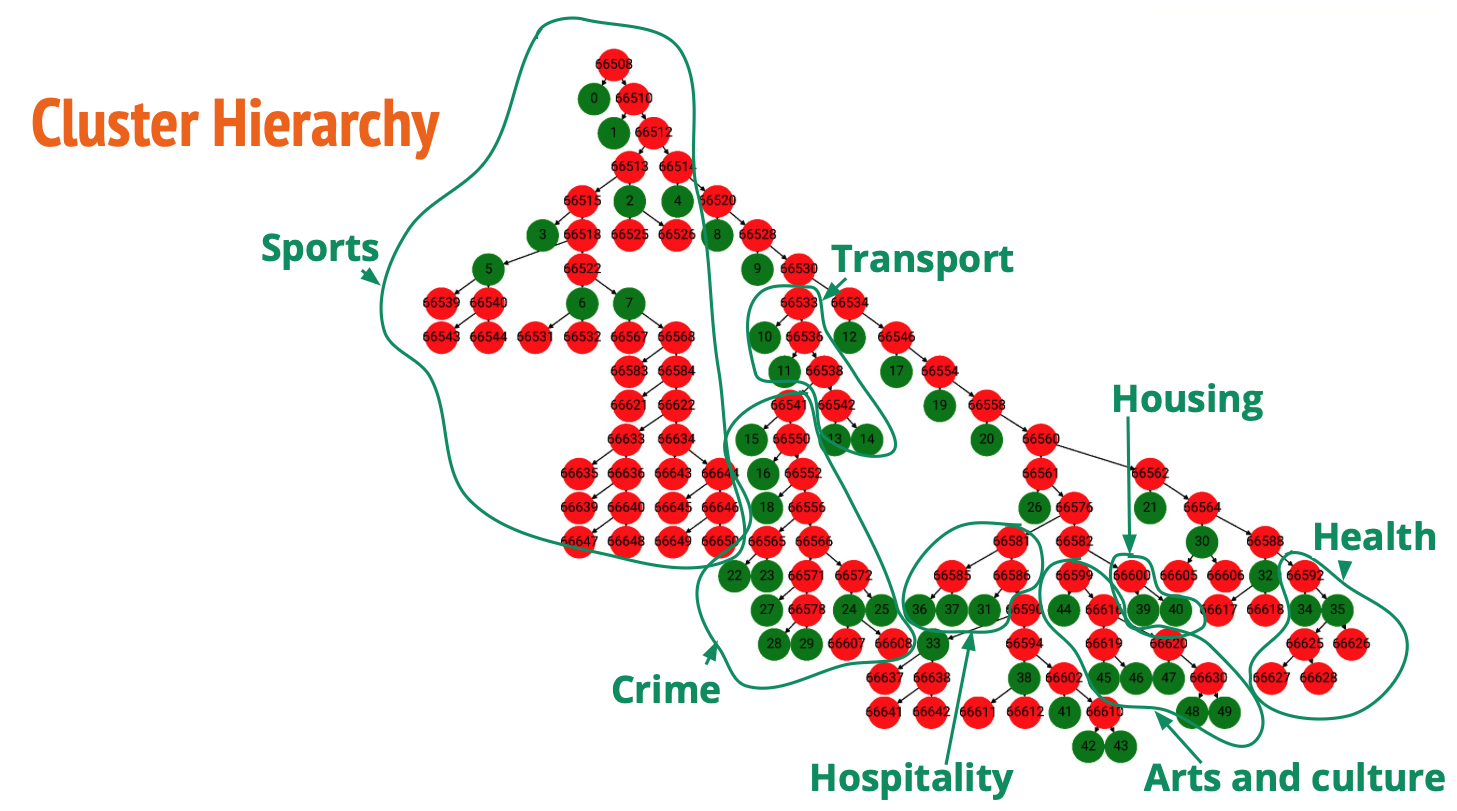}
    \caption{The clustering hierarchy contains semantically meaningful groupings of clusters (themes), which we have annotated in green. Each green node corresponds to a cluster id, while red nodes are candidate clusters that were not selected by the clustering algorithm.}
    \label{fig:clusterhier-groups}
\end{figure}

\paragraph{Verifying the Cluster Hierarchy}
\label{sec:hier}
We now verify that the clustering hierarchy makes sense.
In fact, we see that the themes we specified in the previous section are subtrees in the cluster hierarchy.
As can be seen in~\cref{fig:clusterhier-groups}, in addition to the Sports related topics, we also see clusters related to other themes: Crime, Transport, Hospitality, Arts and Culture, Health and Housing.
An exception is a theme related to business, which appears scattered in the hierarchy.

\subsection{Quantitative Evaluation}
\label{subsec:quant}
As per quantitative evaluation we checked two things:
1) We checked that the cluster scores were capturing known information about neighbourhoods. To do so, we assessed how well our clusters correlate with known features from the $2020$v2 Scottish Index of Multiple Deprivation (SIMD).
2) We verified that the clusters were coherent i.e. similar articles were clustered together. To do so, we annotated pairs of articles as to whether they were on the same topic or not (see~\cref{sec:clusterann}).

\subsubsection{Crime Topics Correlate with SIMD Features}
We used the \emph{crime rate} metric from the SIMD features which measures the crime rate per 10,000 population as recorded by Police Scotland in 2017-2018. Crimes include violence, sexual offences, vandalism, drug offences and common assault, but some crimes have been excluded due to data quality issues or because they are directed at businesses rather than neighbourhoods (e.g. shoplifting). The rate was rounded to the nearest integer and counts for data zones with less than three counts were suppressed.

\begin{figure}[t]
    \centering
    \begin{subfigure}[b]{.19\textwidth}
    \centering
    \captionsetup{justification=centering}
    \includegraphics[width=\textwidth]{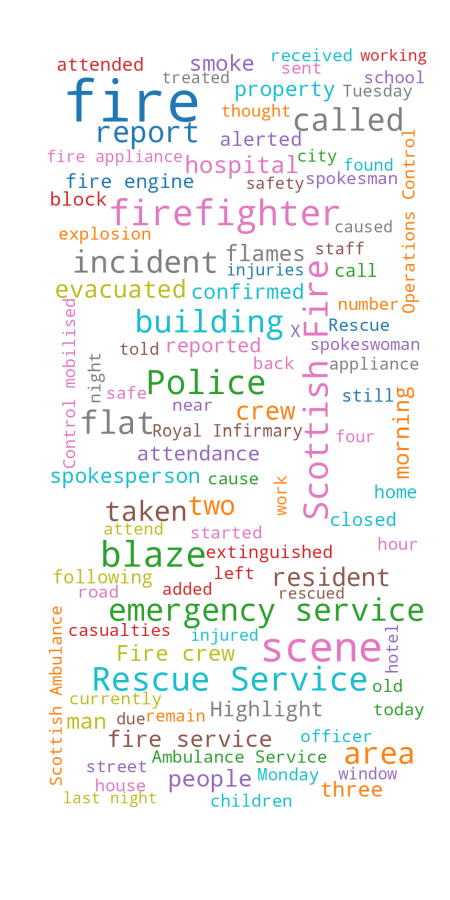}
    \caption{Cluster 27\\$\rho=0.28$}
    \end{subfigure}
    \begin{subfigure}[b]{.19\textwidth}
    \centering
    \captionsetup{justification=centering}
    \includegraphics[width=\textwidth]{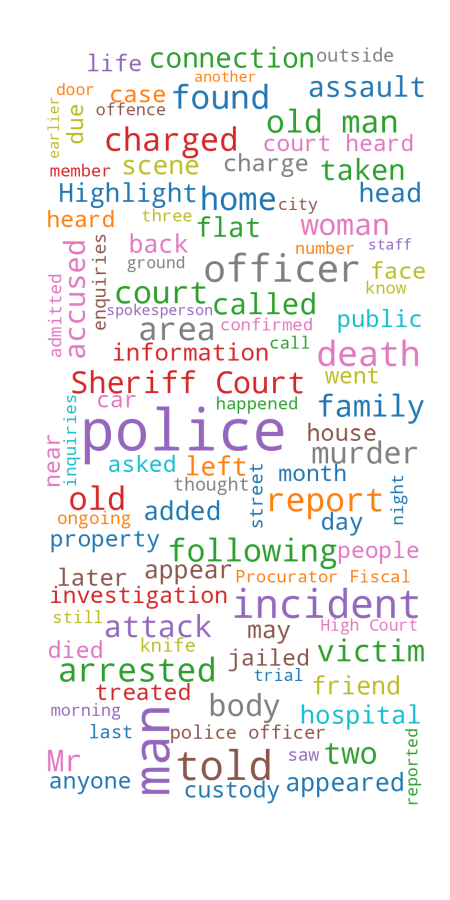}
    \caption{Cluster 24\\$\rho=0.27$}
    \end{subfigure}
    \begin{subfigure}[b]{.19\textwidth}
    \centering
    \captionsetup{justification=centering}
    \includegraphics[width=\textwidth]{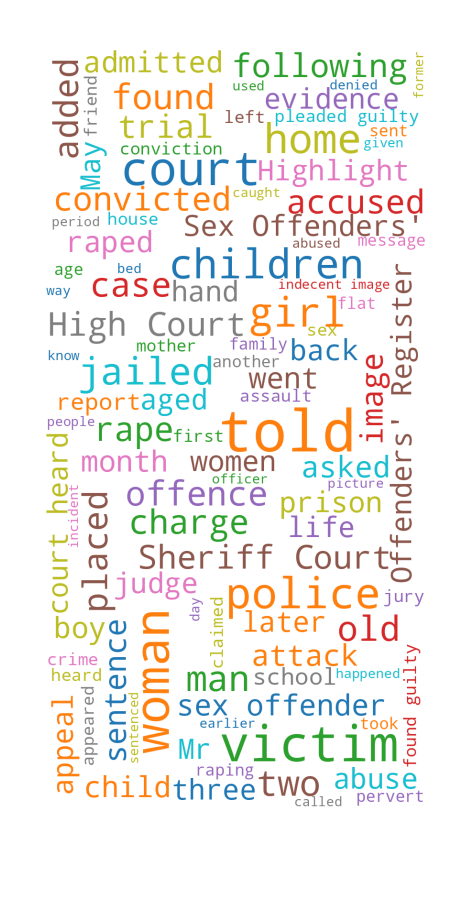}
    \caption{Cluster 25\\$\rho=0.23$}
    \end{subfigure}
    \begin{subfigure}[b]{.19\textwidth}
    \centering
    \captionsetup{justification=centering}
    \includegraphics[width=\textwidth]{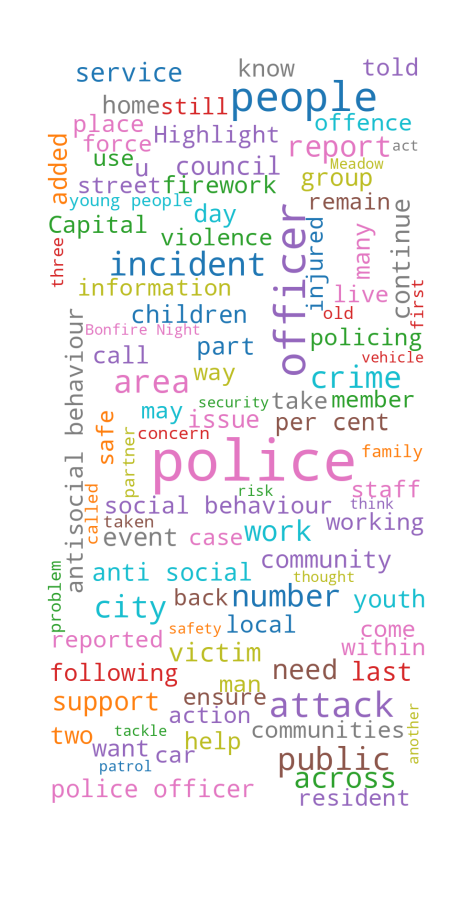}
    \caption{Cluster 16\\$\rho=0.22$}
    \end{subfigure}
    \begin{subfigure}[b]{.19\textwidth}
    \centering
    \captionsetup{justification=centering}
    \includegraphics[width=\textwidth]{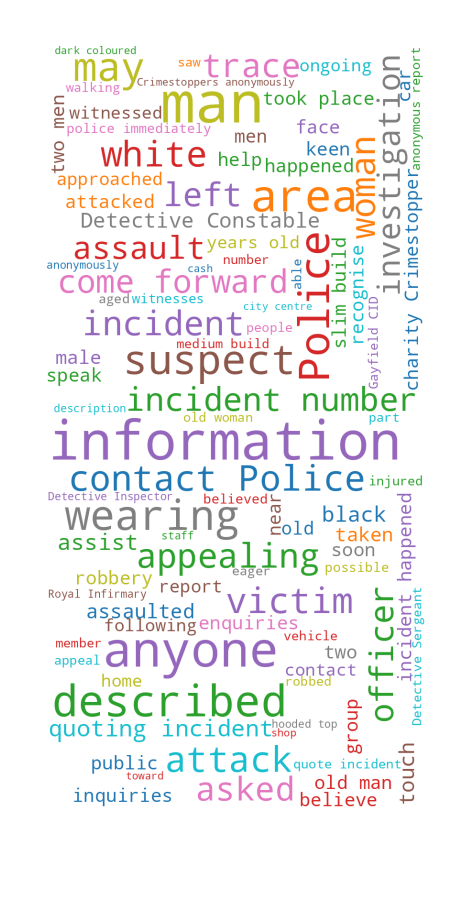}
    \caption{Cluster 22\\$\rho=0.21$}
    \end{subfigure}
    \caption{Word clouds of clusters (top 5, ordered left to right) which are most correlated to SIMD 2020v2 crime rate metric, measured by Spearman's rho, $\rho$.}
    \label{fig:cluster-wc-simd}
\end{figure}

We evaluate our cluster scores by measuring the correlation between each data zone's cluster scores to the crime rate from SIMD. We use Spearman's rho ($\rho$) which is more suitable for comparing measurements that may not be on the same scale or units since it looks at relative rankings instead of magnitudes. We obtain a Spearman's rho correlation per cluster, which can be between -1 and 1, with 1 meaning perfect correlation, 0 meaning no correlation and -1 meaning negative correlation. For our case, Spearman's rho captures whether ranking the data zones in increasing order using their corresponding cluster score is similar to the ranking obtained if we rank the data zones using the SIMD crime rate. As such, clusters that are capturing some aspects related to crime should have a higher correlation.

In~\cref{fig:cluster-wc-simd} we list the clusters with the largest correlations and their corresponding word clouds.
As can be seen, the clusters that surface as having the largest correlations are indeed related to crime.
In addition, as we will see in~\cref{fig:clusterhier-groups}, these clusters are all grouped together in the cluster hierarchy extracted by HDBScan, further supporting the argument that crime data can be reasonably captured for various areas. We conclude that news, and therefore our clusters, indeed capture neighbourhood information related to crime. %
However, one caveat with our analysis is that the correlations are not very large in magnitude.
A plausible explanation for this is that newspapers focus on particular types of crimes in particular places, and this leads to somewhat different spatial patterns to crime in the SIMD. Furthermore, newspapers potentially pick up crime that is not reported to Police (e.g. low level things like flytipping), and as such is not in the SIMD statistics.

\subsubsection{Article Pairs are Mostly Clustered Correctly}

In order to verify the clusters, we check whether similar pairs of articles end up in the same cluster.
To this end, we use the annotated dataset we described in~\cref{sec:clusterann}.
We obtain a Macro-F1 score of $78\%$, which is high given the caveats we will discuss in the Error Analysis section.

\paragraph{Evaluation Metric}
\label{sec:cluster-eval}
Given the output clusters, we compute a Macro-F1 score based on the TP, TN, FP and FN pairs.
This metric is focussed on local coherence, but may be missing characteristics of global coherence of the clusters.~\footnote{Our extracted clusters and annotations can be found at~\href{https://docs.google.com/spreadsheets/d/10A4QIsUmZC9XRqnYzHtuzbY83gvOOgsKipS8Mn\_vk84}{here}. We used the masked data zone analysis.}

\paragraph{Results}
\label{sec:results}
We evaluated our clustering results using the annotated article pairs.
During hyperparameter selection, we varied the tf-idf vocabulary size $|V|$, and UMAP hyperparameters which include the dimensionality of the projection $d$ and the number of UMAP neighbours $n$.
We obtained 51 clusters and a Macro-F1 score of 78\% with the best setup, which used $|V|=20000$, $n=5$ and $d=10$.
See~\ref{app:cross} for the full details.

\paragraph{Dealing with HDBScan Outliers}
Annotating pairs of articles as to whether they should belong to the same cluster or not comes with one caveat: what do we do with the case of outliers, i.e.~the articles to which HDBScan assigns the -1 cluster?\footnote{We abuse the term cluster here, since these articles may be scattered across space and need not be cohesive.} For this study we did the simplest thing possible and treated the outliers as another cluster. An alternative approach would be to ignore article pairs when both are assigned -1, treating the -1 cluster as a ``do not predict'' case. However, this approach is not ideal either, since then we are encouraging clustering outcomes that assign -1 to all hard clustering decisions.

\paragraph{Error Analysis}
We now further analyse how our annotated article pairs were clustered when using the best choice of hyperparameters, which we mentioned above. For the analysis, we partition the annotated article pairs into 3 categories depending on the clustering predictions. The breakdown of the 700 article pairs into these categories is as follows: 40 pairs were classified as outliers, 195 pairs were assigned to the same cluster (but were not outliers), and 465 pairs were assigned to different clusters. We further analyse these cases below as to how they were annotated.

\paragraph{Article Pairs assigned as Outliers (40)}
Herein we quantify the impact of outliers on our evaluation.
Out of the 40 article pairs in this category, 23 were annotated as \emph{Not related}, 4 as \emph{Vaguely related}, 4 as \emph{Somewhat related} and 9 as \emph{Very related}, see~\cref{tab:outliers} for more details.
All 9 \emph{Very related} article pairs were indeed related (i.e.~there was no annotation error), and included granular themes like roadworks, fishing, the Edinburgh Zoo, funding the arts and cinema news. Although including the above articles in the evaluation can lower the evaluation scores (e.g.~we made 23 mistakes), we see that this is a non-trivial problem: alternative choices for evaluation, such as ignoring these articles, would also miss out nuance in some way.

\paragraph{Article Pairs assigned to same Clusters (195)}
As expected, most of these articles were annotated as Very related.
More specifically, 147 pairs were annotated as \emph{Very related},
 33 as \emph{Somewhat related}, 9 as \emph{Vaguely related},
and 6 pairs were annotated as \emph{Not related}.
We inspected these 6 article pairs~\footnote{These article pairs are highlighted in purple~\href{https://docs.google.com/spreadsheets/d/1W9UIvSBsEJPLqajr8\_CCtFgjqddk_mqVjBJClGVBGdQ/edit\#gid=1019576175}{here}.} and all 6 were incorrectly labelled as \textit{Not Related}. The errors were due to only relying on the title to judge relatedness. In two cases the title did not capture what was in the text. For example, one article pair were editorials by MPs that spanned multiple overlapping subjects. Another was about a film on the mental health struggles during the pandemic. The first article described the film, and the description was very similar to the second article, which is a retrospective on the pandemic one year after the first lockdown.
In the remaining 4 articles, the annotator misinterpreted the article or did not have enough background knowledge to understand the connection: e.g.~one pair of articles on the Edinburgh fringe and vertical farming turned out to contain similar content, since the vertical farming article was actually on the harvest festival, which involves music and performances and as such is not too different from the fringe festival.

\paragraph{Article Pairs assigned to different Clusters (465)}
Out of the 465 article pairs, 285 were annotated as \emph{Not related}, 64 as \emph{Vaguely related}, 62 as \emph{Somewhat related} and 54 as \emph{Very related}. The reason for so many articles (54) annotated as \emph{Very related} while being assigned to different clusters is that a) for 24 of the pairs one of the articles is classified as an outlier (-1 cluster) and b) for the remaining pairs, clusters are actually very related, i.e.~out of the 30 remaining pairs 17 are assigned to the same theme using the cluster hierarchy seen in~\cref{fig:hierarchy}.

\section{Discussion and Limitations}\label{sec:discussion}

In this paper we have proposed a simple methodology for characterising neighbourhoods based on clustering geoparsed news articles.
Furthermore, we have shown via a confluence of evidence, both from a qualitative (~\cref{subsec:qual}) and a quantitative analysis (~\cref{subsec:quant}) that the clusters we have extracted from news articles are sensible and reflect some characteristics of the real world. Albeit, these characteristics are skewed via the lens of news.
We posed a research question of whether we can extract information from ephemeral data that relates to communities.
We find that community data is available and can permit further research to link this to resilience in neighbourhoods and elucidate why some neighbourhoods have better health outcomes than expected. We have shown it is possible to characterise neighbourhoods and data zones.  The findings are promising for enabling new epidemiological research on place effects on health. One additional finding is that we expect our results for neighbourhoods to be more robust than those for more fine grained zones, as we explain below.

\subsection{Characterisations are more robust on the neighbourhood level}
Transforming the clustering results into a characterisation for fine grained neighbourhoods is challenging because:

\paragraph{a) We have much less data for granular areas}
Location mentions have a Zipfian distribution, with densely populated areas and coarse areas mentioned much more than fine grained and sparsely populated areas. As such, our estimates for fine grained areas are always going to have higher variance.

\paragraph{b) Geoparsing is especially noisy for fine grained locations}
Mention of place in news text and the mappings we use for geoparsing are not always consistent with fine grained data zones. This is particularly apparent with the names of large neighbourhoods (e.g.\ Portobello). When a news report uses the name Portobello it usually has the sense of the entire neighbourhood, but this covers six fine-grained data zones in the SIMD. Our georesolution method, which relies on Ordance Survey point data, puts ``Portobello" in one particular data zone, which then appears much more talked about than the other Portobello data zones. 

Similarly, while short streets (e.g.\ Buccleuch Place) will be entirely contained within one data zone, other longer ones (e.g.\ Queensferry Road) will span many data zones. Our georesolution method maps to point coordinates, so some streets will be disproportionately (and often incorrectly) mapped to one particular data zone. 

A further example in this vein concerns buildings such as schools, which are frequently mentioned in news articles. For example, Abbeyhill Primary School falls within a single fine grained data zone but it serves a catchment area that includes multiple data zones. For this reason, the larger neighbourhood is possibly more pertinent than one particular data zone.

We conclude that we can more robustly characterise broader regions than data zones, as data zones are too fine grained.
While our conclusion is based on our data, the Zipfian distribution of location mentions is not specific to our dataset.
As such, our conclusion is likely to be true for many datasets.

\section{Future Work}
In this Section we highlight a few possible directions for future work.
\label{sec:future}

\paragraph{Further Verification of Results}
While it is important that we verify our outputs by checking that they agree with official statistics of neighbourhoods, we can only do so for statistics that are already available.
An important next step is to test the validity of the area-level measures by examining associations against health outcomes with a biologically plausible pathway.

\paragraph{Neighbourhood Change over Time}
In this paper we worked with limited data, both in size (only 66k articles), type (newspaper articles) and time (5 years). Our promising results suggest that we may gain further insights if we work with data over longer time spans, this would also enable investigating how neighbourhoods change over time, as done by~\citep{bechini2022news}.

\paragraph{Improved Modelling for Locations}
Our results clearly indicate that locations that are mentioned in fewer articles have noisier characterisations. In this paper we explained why solving this is hard, and resorted instead to characterising more granular mentions for which we have aggregate statistics.
At the same time, if we had more data we could also
try to smooth the statistics of neighbourhoods by leveraging the adjacency structure of data zones.
For example, this problem could be framed as node feature prediction using a Graph Neural Network~\citep{Zhu2020}.
Additional future directions include further investigation of the clustering hyperparameters, improving feature extraction and using a probabilistic model for characterising neighbourhoods.

\section{Ethics}

Ethics approval for this work on ``Artificial Intelligence and Multimorbidity: Clustering in Individuals, Space and Clinical Context (AIM-CISC), Objective 3'' (ID: 6f8d-388d-ed15-9b4c)  was obtained on 15/07/2021.

\subsection{Challenges arising from the data}
Views expressed in news can be biased and are generally non-representative of the whole population. This should be taken into account when considering potential uses of the data and systems developed from it.

\subsection{Challenges arising from quantitative analyses}
Quantitative analyses of locations and neighbourhoods need to be thoroughly vetted for errors and omissions and cross-validated with other data sources.
In this article, we took a first step and cross-validated our cluster metrics with crime statistics from SIMD, but further validation would be beneficial.
We have highlighted the strengths and limitations of our analysis to help guide the use and further development of such approaches.

\section{Funding}
This work was funded by the National Institute for Health Research (NIHR) Artificial Intelligence and Multimorbidity: Clustering in Individuals, Space and Clinical Context (AIM-CISC) grant NIHR202639. AM is also funded by the REALITIES in Health Disparities project, funded by UK Research and Innovations and led by the Arts Humanities Research Council.
BA and AM are also funded by Legal and General Group as part of their corporate social responsibility (CSR) programme, providing a research grant to establish the independent Advanced Care Research Centre at the University of Edinburgh. The funder had no role in the conduct of the study, interpretation or the decision to submit for publication. The views expressed are those of the authors and not necessarily those of the NIHR, the Department of Health and Social Care or Legal and General.

\bibliographystyle{elsarticle-harv} 
\bibliography{bib}

\appendix

\section{Word Clouds}
\label{app:wclouds}
\begin{figure}[t]
    \centering
    \begin{subfigure}[b]{.19\textwidth}
    \centering
    \captionsetup{justification=centering}
    \includegraphics[width=\textwidth]{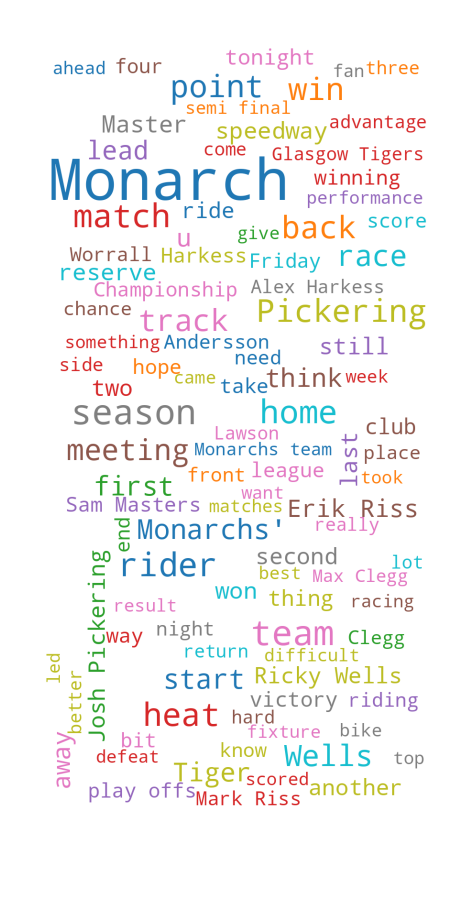}
    \vspace{-2.25\baselineskip}
    \caption{Cluster 0: ``Motorbiking''}
    \end{subfigure}
    \begin{subfigure}[b]{.19\textwidth}
    \centering
    \captionsetup{justification=centering}
    \includegraphics[width=\textwidth]{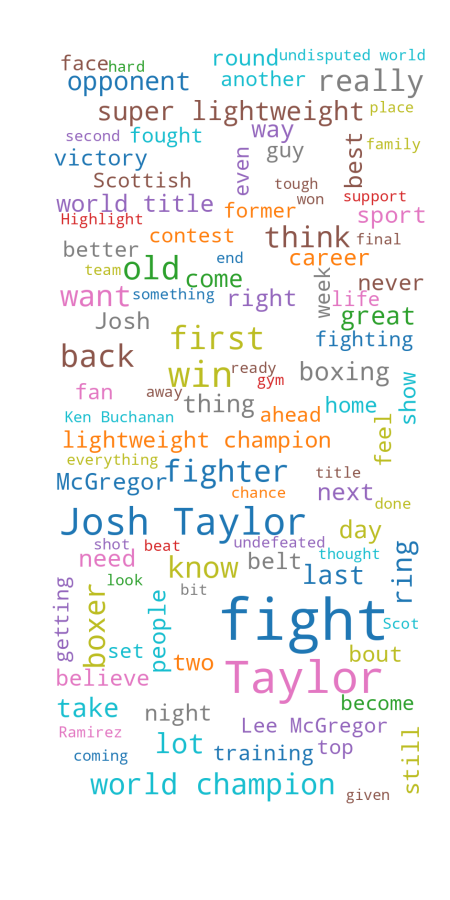}
    \vspace{-2.25\baselineskip}
    \caption{Cluster 1: ``Boxing''}
    \end{subfigure}
    \begin{subfigure}[b]{.19\textwidth}
    \centering
    \captionsetup{justification=centering}
    \includegraphics[width=\textwidth]{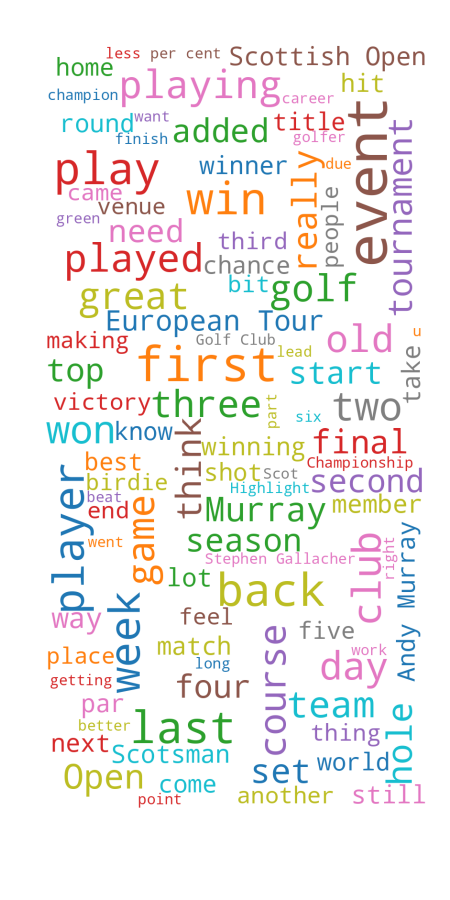}
    \vspace{-2.25\baselineskip}
    \caption{Cluster 2: ``Tennis''}
    \end{subfigure}
    \begin{subfigure}[b]{.19\textwidth}
    \centering
    \captionsetup{justification=centering}
    \includegraphics[width=\textwidth]{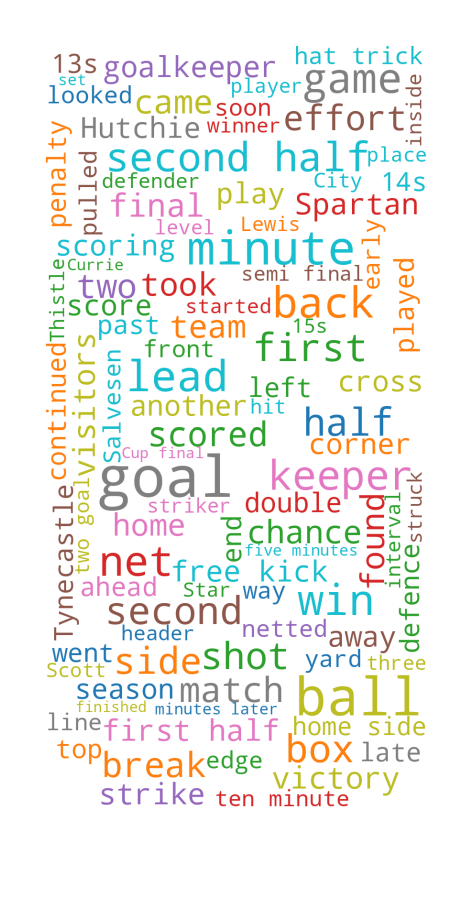}
    \vspace{-2.25\baselineskip}
    \caption{Cluster 3: ``Football''}
    \end{subfigure}
    \begin{subfigure}[b]{.19\textwidth}
    \centering
    \captionsetup{justification=centering}
    \includegraphics[width=\textwidth]{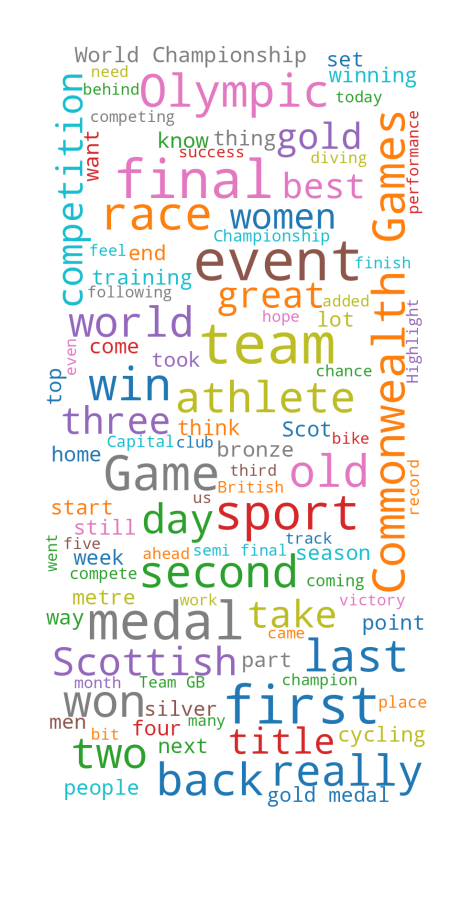}
    \vspace{-2.25\baselineskip}
    \caption{Cluster 4: ``Sporting events''}
    \end{subfigure}
    \begin{subfigure}[b]{.19\textwidth}
    \centering
    \captionsetup{justification=centering}
    \includegraphics[width=\textwidth]{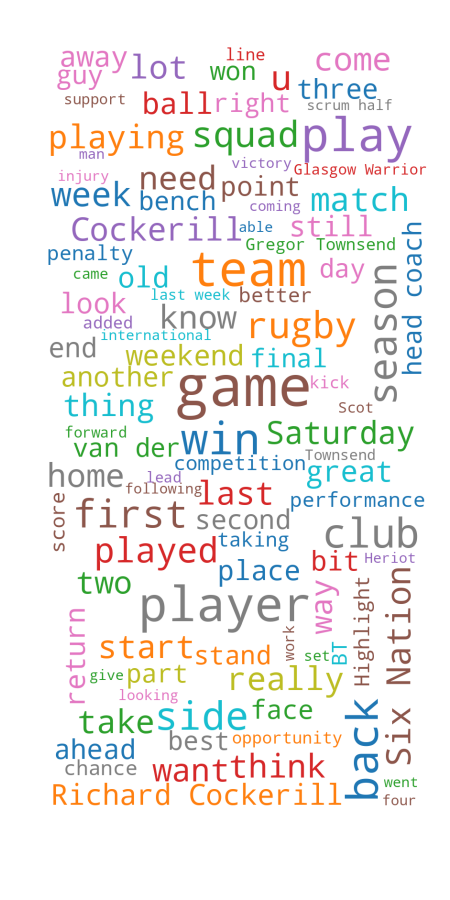}
    \vspace{-2.25\baselineskip}
    \caption{Cluster 5: ``Rugby''}
    \end{subfigure}
    \begin{subfigure}[b]{.19\textwidth}
    \centering
    \captionsetup{justification=centering}
    \includegraphics[width=\textwidth]{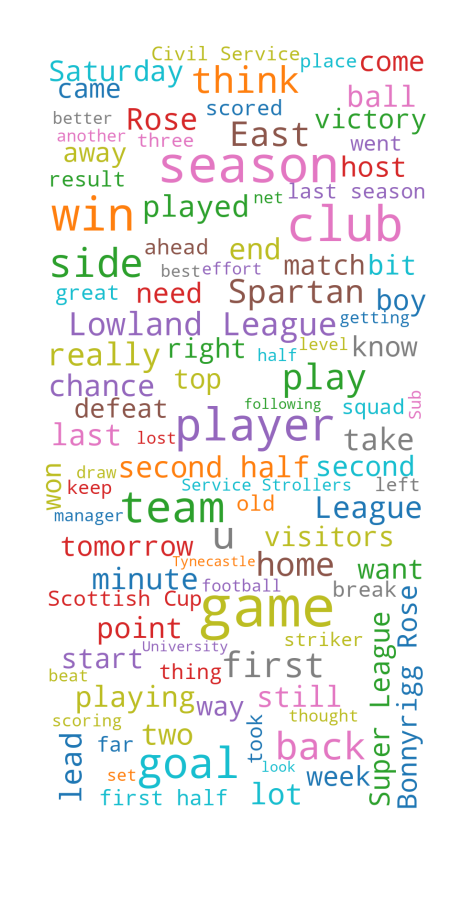}
    \vspace{-2.25\baselineskip}
    \caption{Cluster 6: ``Football''}
    \end{subfigure}
    \begin{subfigure}[b]{.19\textwidth}
    \centering
    \captionsetup{justification=centering}
    \includegraphics[width=\textwidth]{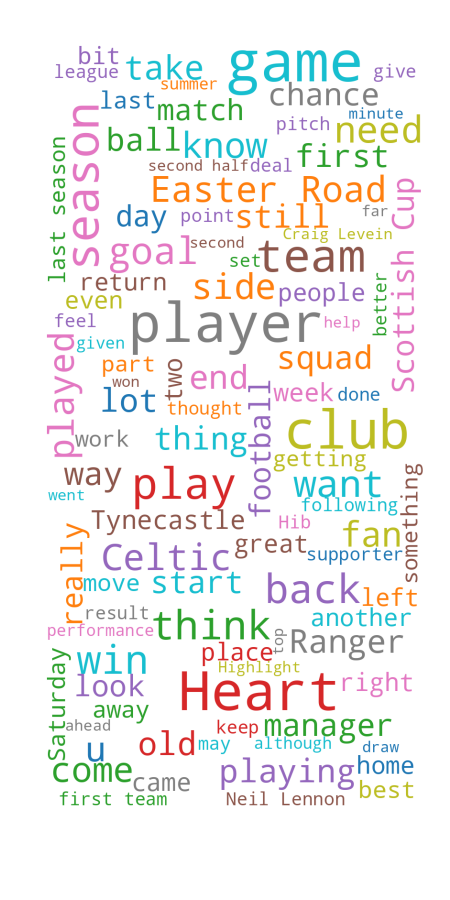}
    \vspace{-2.25\baselineskip}
    \caption{Cluster 7: ``Football''}
    \end{subfigure}
    \begin{subfigure}[b]{.19\textwidth}
    \centering
    \captionsetup{justification=centering}
    \includegraphics[width=\textwidth]{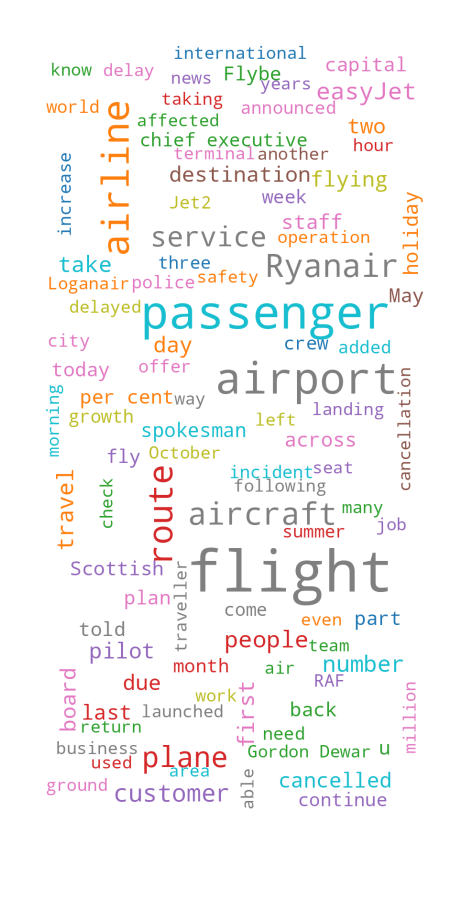}
    \vspace{-2.25\baselineskip}
    \caption{Cluster 8: ``Air travel''}
    \end{subfigure}
    \begin{subfigure}[b]{.19\textwidth}
    \centering
    \captionsetup{justification=centering}
    \includegraphics[width=\textwidth]{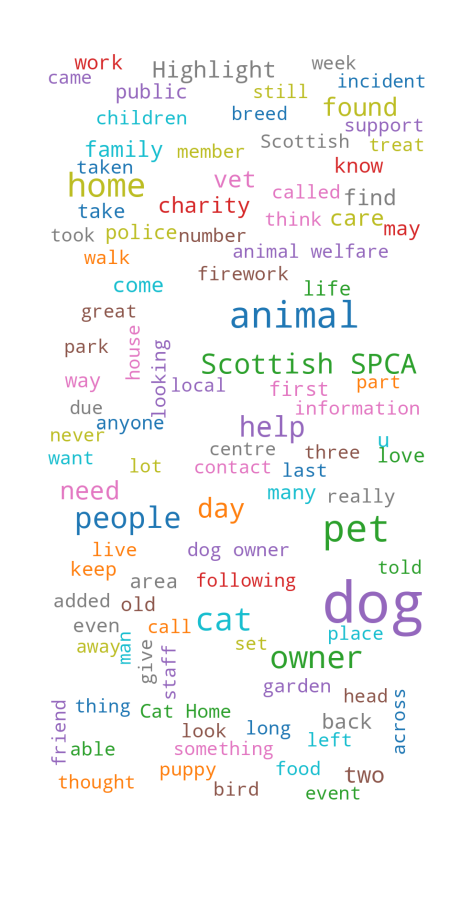}
    \vspace{-2.25\baselineskip}
    \caption{Cluster 9: ``Animals''}
    \end{subfigure}
    \begin{subfigure}[b]{.19\textwidth}
    \centering
    \captionsetup{justification=centering}
    \includegraphics[width=\textwidth]{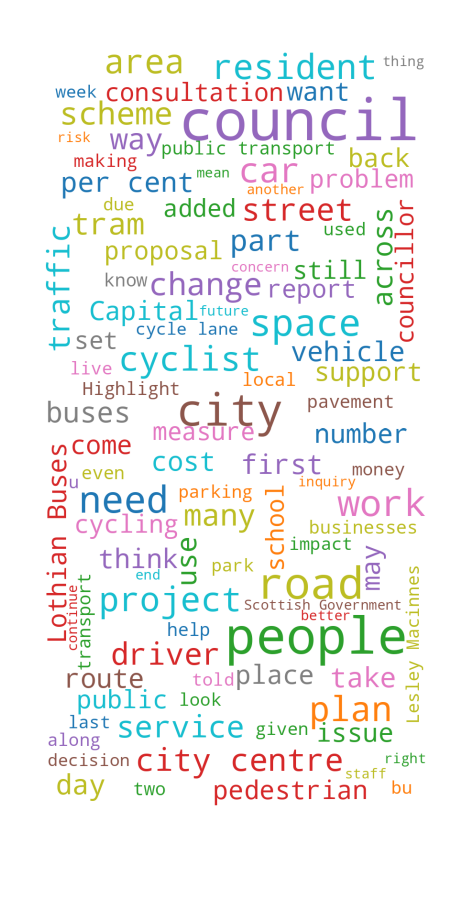}
    \vspace{-2.25\baselineskip}
    \caption{Cluster 10: ``City transport''}
    \end{subfigure}
    \begin{subfigure}[b]{.19\textwidth}
    \centering
    \captionsetup{justification=centering}
    \includegraphics[width=\textwidth]{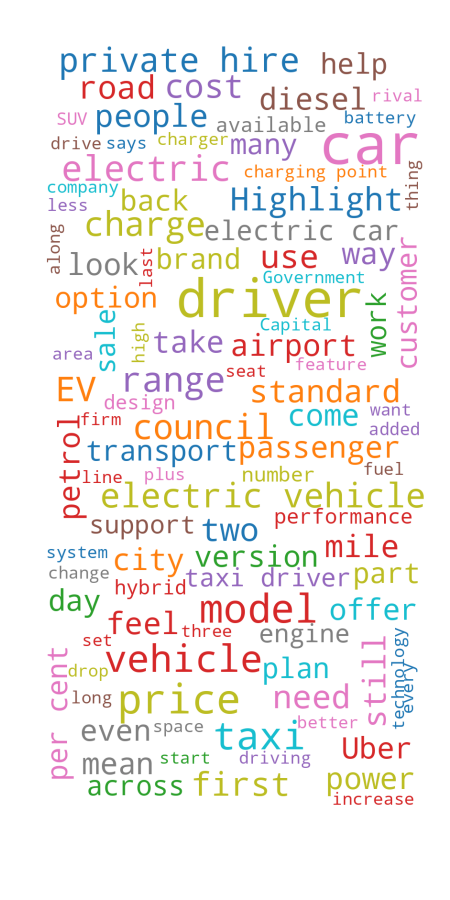}
    \vspace{-2.25\baselineskip}
    \caption{Cluster 11: ``Car transport''}
    \end{subfigure}
    \begin{subfigure}[b]{.19\textwidth}
    \centering
    \captionsetup{justification=centering}
    \includegraphics[width=\textwidth]{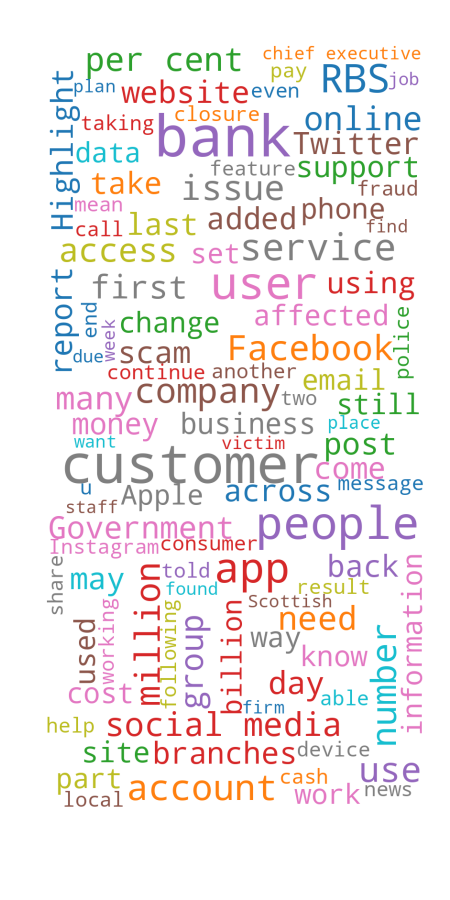}
    \vspace{-2.25\baselineskip}
    \caption{Cluster 12: ``Banking''}
    \end{subfigure}
    \begin{subfigure}[b]{.19\textwidth}
    \centering
    \captionsetup{justification=centering}
    \includegraphics[width=\textwidth]{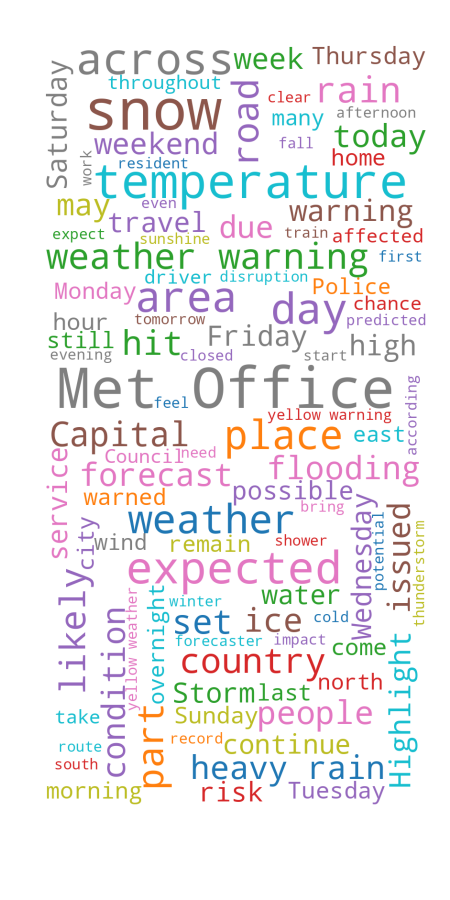}
    \vspace{-2.25\baselineskip}
    \caption{Cluster 13: ``Weather''}
    \end{subfigure}
    \begin{subfigure}[b]{.19\textwidth}
    \centering
    \captionsetup{justification=centering}
    \includegraphics[width=\textwidth]{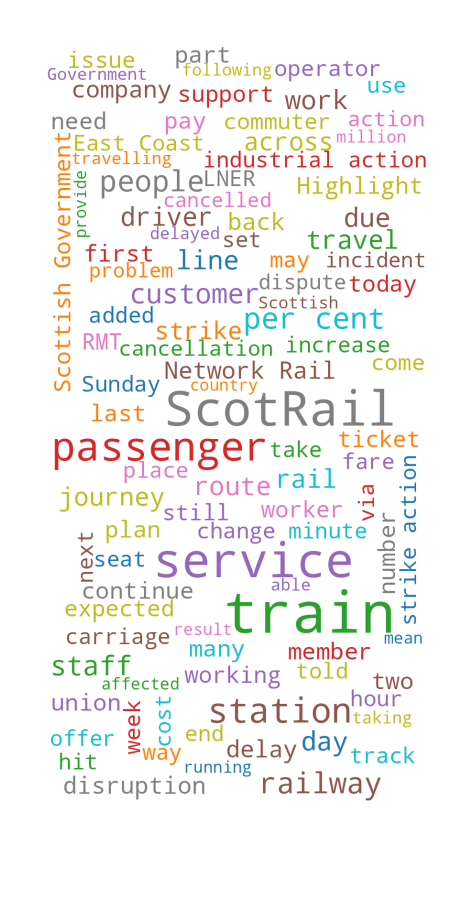}
    \vspace{-2.25\baselineskip}
    \caption{Cluster 14: ``Train transport''}
    \end{subfigure}
    \label{fig:clusters-1}
\end{figure}

\begin{figure}[t]
    \centering
    \begin{subfigure}[b]{.19\textwidth}
    \centering
    \captionsetup{justification=centering}
    \includegraphics[width=\textwidth]{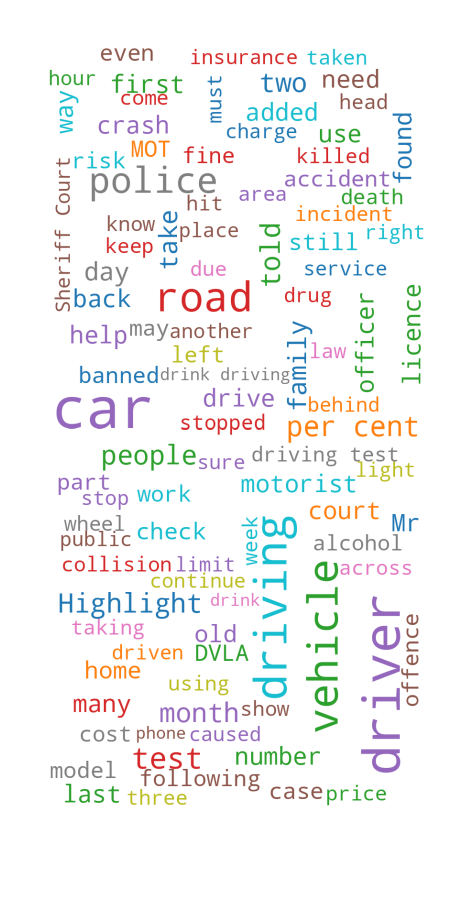}
    \vspace{-2.25\baselineskip}
    \caption{Cluster 15: ``Driving incidents''}
    \end{subfigure}
    \begin{subfigure}[b]{.19\textwidth}
    \centering
    \captionsetup{justification=centering}
    \includegraphics[width=\textwidth]{images/cluster_16.pdf}
    \vspace{-2.25\baselineskip}
    \caption{Cluster 16: ``Police / crime''}
    \end{subfigure}
    \begin{subfigure}[b]{.19\textwidth}
    \centering
    \captionsetup{justification=centering}
    \includegraphics[width=\textwidth]{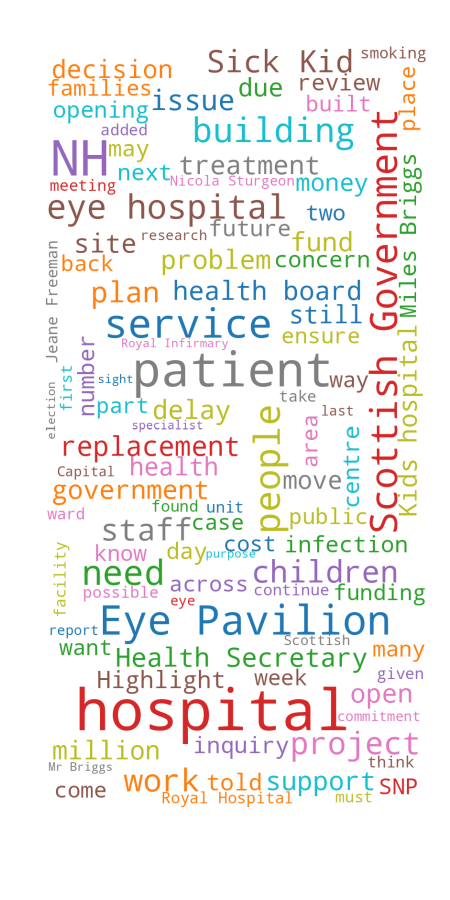}
    \vspace{-2.25\baselineskip}
    \caption{Cluster 17: ``Healthcare''}
    \end{subfigure}
    \begin{subfigure}[b]{.19\textwidth}
    \centering
    \captionsetup{justification=centering}
    \includegraphics[width=\textwidth]{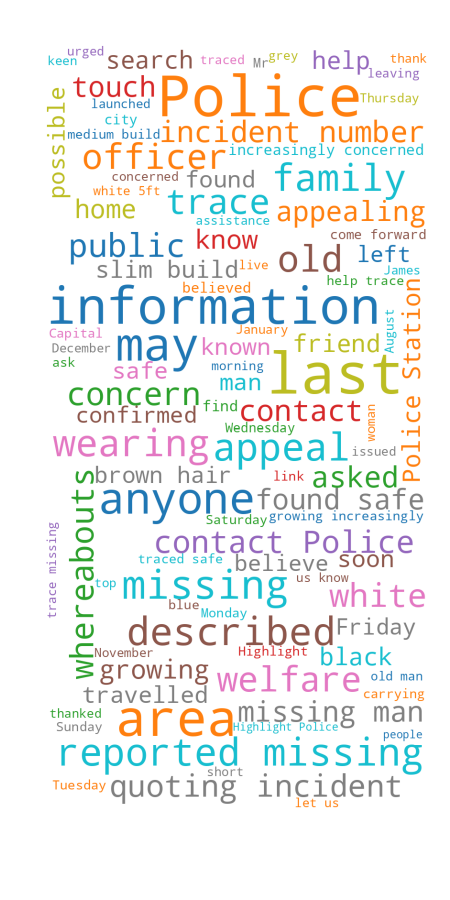}
    \vspace{-2.25\baselineskip}
    \caption{Cluster 18: ``Missing persons''}
    \end{subfigure}
    \begin{subfigure}[b]{.19\textwidth}
    \centering
    \captionsetup{justification=centering}
    \includegraphics[width=\textwidth]{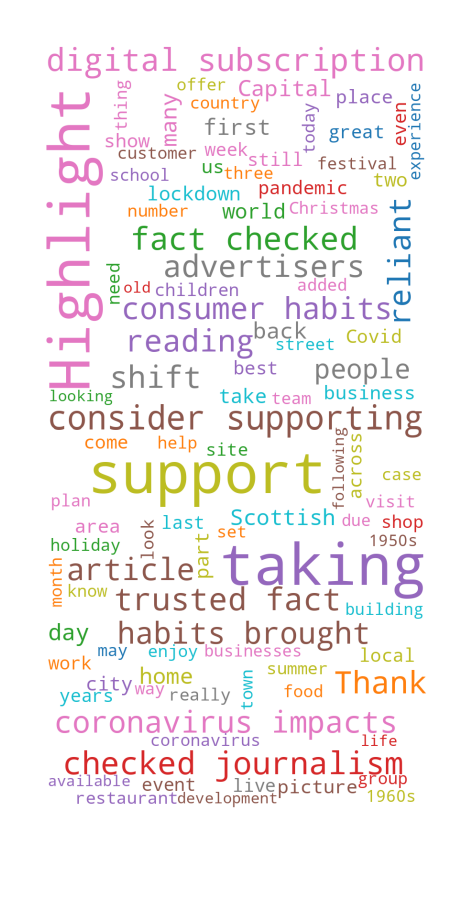}
    \vspace{-2.25\baselineskip}
    \caption{Cluster 19: ``News paper ads''}
    \end{subfigure}
    \begin{subfigure}[b]{.19\textwidth}
    \centering
    \captionsetup{justification=centering}
    \includegraphics[width=\textwidth]{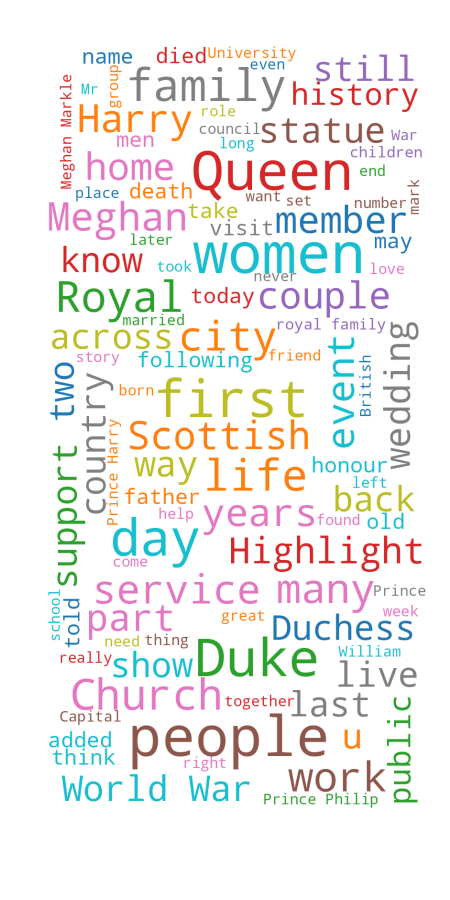}
    \vspace{-2.25\baselineskip}
    \caption{Cluster 20: ``Royal family''}
    \end{subfigure}
    \begin{subfigure}[b]{.19\textwidth}
    \centering
    \captionsetup{justification=centering}
    \includegraphics[width=\textwidth]{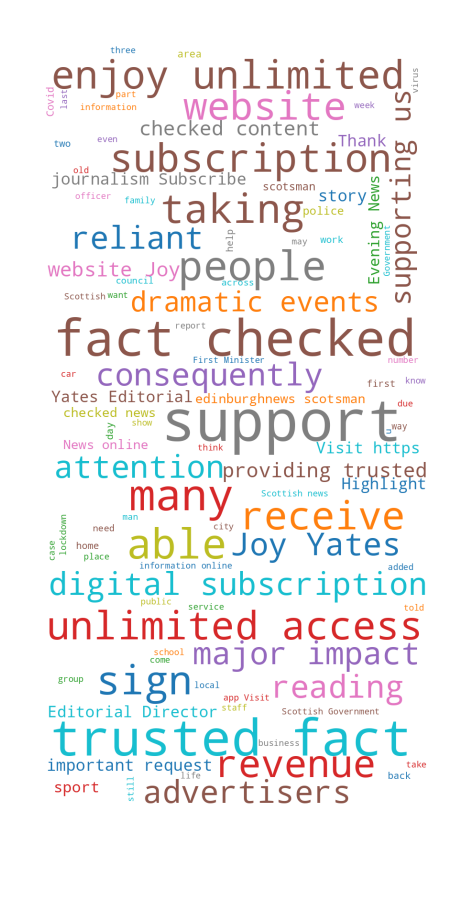}
    \vspace{-2.25\baselineskip}
    \caption{Cluster 21: ``Newspaper ads''}
    \end{subfigure}
    \begin{subfigure}[b]{.19\textwidth}
    \centering
    \captionsetup{justification=centering}
    \includegraphics[width=\textwidth]{images/cluster_22.pdf}
    \vspace{-2.25\baselineskip}
    \caption{Cluster 22: ``Crime incidents''}
    \end{subfigure}
    \begin{subfigure}[b]{.19\textwidth}
    \centering
    \captionsetup{justification=centering}
    \includegraphics[width=\textwidth]{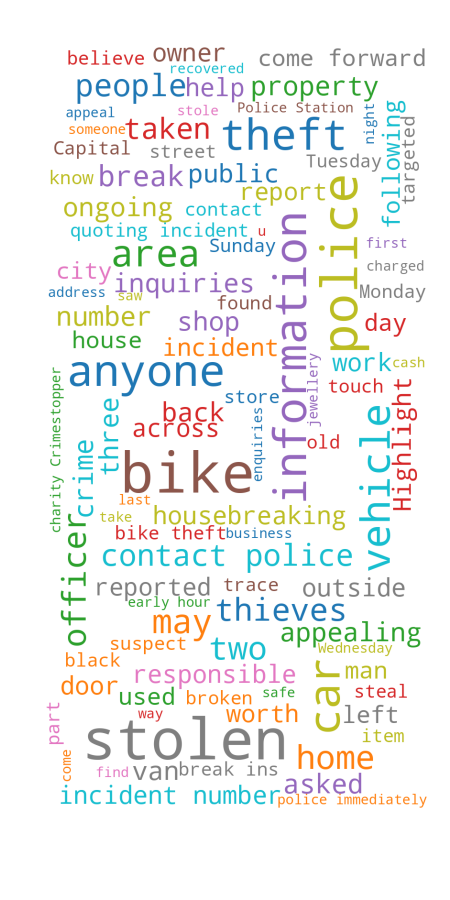}
    \vspace{-2.25\baselineskip}
    \caption{Cluster 23: ``Thefts''}
    \end{subfigure}
    \begin{subfigure}[b]{.19\textwidth}
    \centering
    \captionsetup{justification=centering}
    \includegraphics[width=\textwidth]{images/cluster_24.pdf}
    \vspace{-2.25\baselineskip}
    \caption{Cluster 24: ``Crime incidents''}
    \end{subfigure}
    \begin{subfigure}[b]{.19\textwidth}
    \centering
    \captionsetup{justification=centering}
    \includegraphics[width=\textwidth]{images/cluster_25.pdf}
    \vspace{-2.25\baselineskip}
    \caption{Cluster 25: ``Sex offences''}
    \end{subfigure}
    \begin{subfigure}[b]{.19\textwidth}
    \centering
    \captionsetup{justification=centering}
    \includegraphics[width=\textwidth]{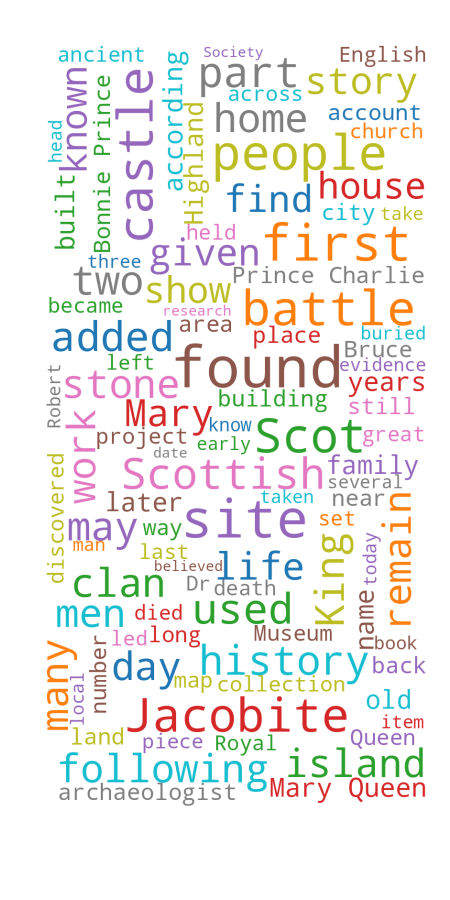}
    \vspace{-2.25\baselineskip}
    \caption{Cluster 26: ``History''}
    \end{subfigure}
    \begin{subfigure}[b]{.19\textwidth}
    \centering
    \captionsetup{justification=centering}
    \includegraphics[width=\textwidth]{images/cluster_27.pdf}
    \vspace{-2.25\baselineskip}
    \caption{Cluster 27: ``Fire incidents''}
    \end{subfigure}
    \begin{subfigure}[b]{.19\textwidth}
    \centering
    \captionsetup{justification=centering}
    \includegraphics[width=\textwidth]{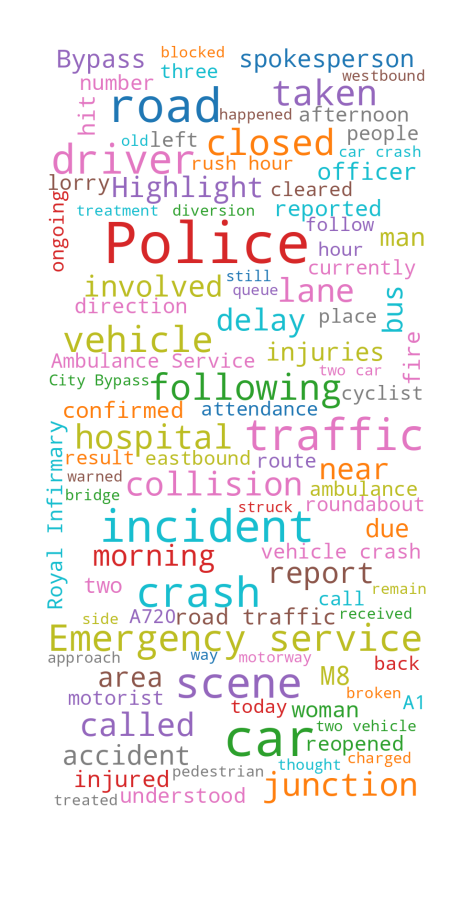}
    \vspace{-2.25\baselineskip}
    \caption{Cluster 28: ``Road accidents''}
    \end{subfigure}
    \begin{subfigure}[b]{.19\textwidth}
    \centering
    \captionsetup{justification=centering}
    \includegraphics[width=\textwidth]{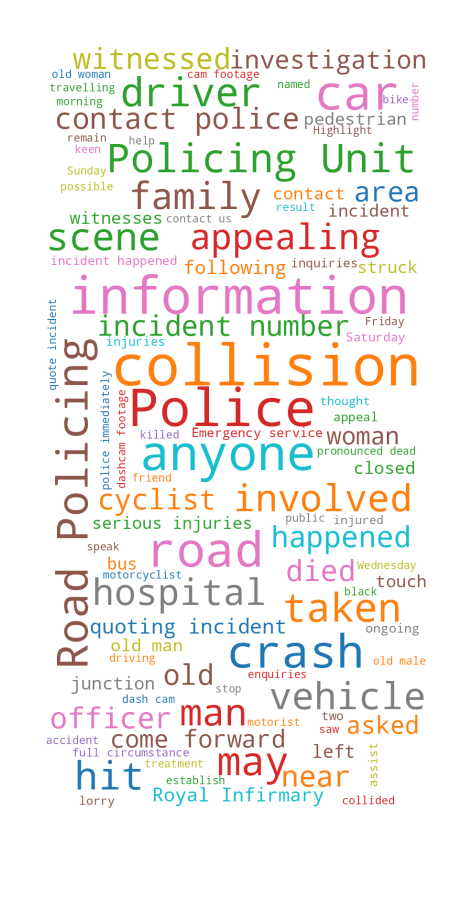}
    \vspace{-2.25\baselineskip}
    \caption{Cluster 29: ``Road accidents''}
    \end{subfigure}
    \label{fig:clusters-2}
\end{figure}

\begin{figure}[t]
    \centering
    \begin{subfigure}[b]{.19\textwidth}
    \centering
    \captionsetup{justification=centering}
    \includegraphics[width=\textwidth]{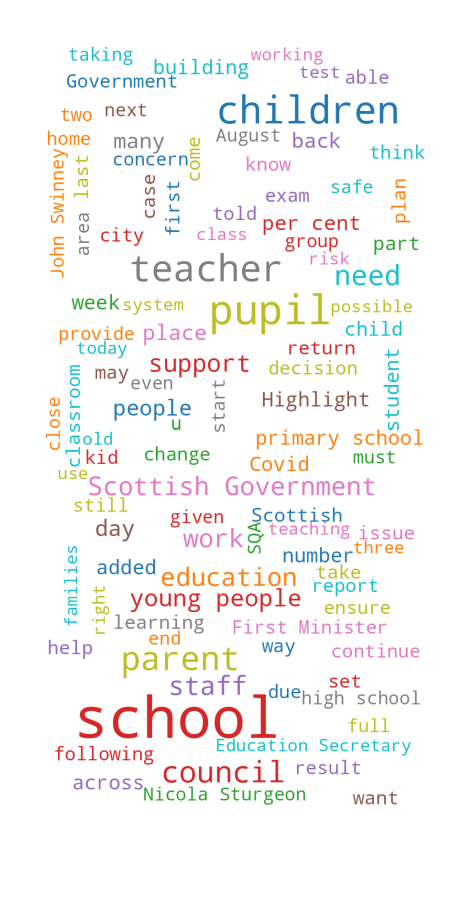}
    \vspace{-2.25\baselineskip}
    \caption{Cluster 30: ``Education''}
    \end{subfigure}
    \begin{subfigure}[b]{.19\textwidth}
    \centering
    \captionsetup{justification=centering}
    \includegraphics[width=\textwidth]{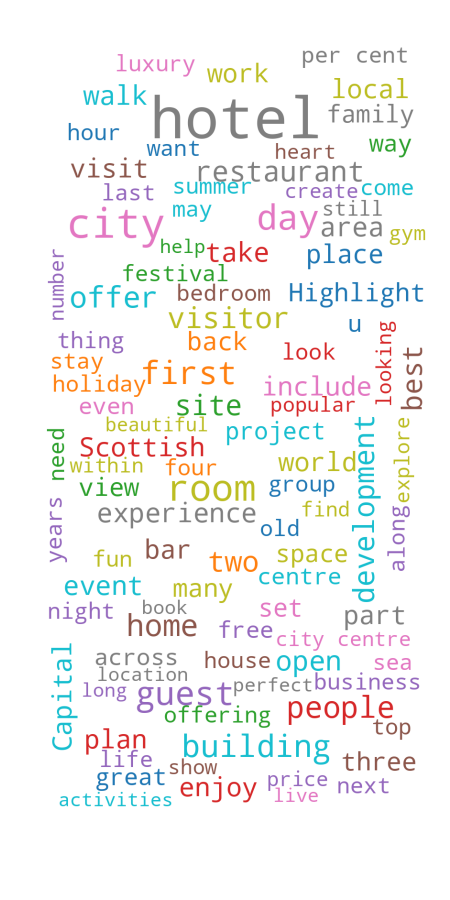}
    \vspace{-2.25\baselineskip}
    \caption{Cluster 31: ``Accommodation''}
    \end{subfigure}
    \begin{subfigure}[b]{.19\textwidth}
    \centering
    \captionsetup{justification=centering}
    \includegraphics[width=\textwidth]{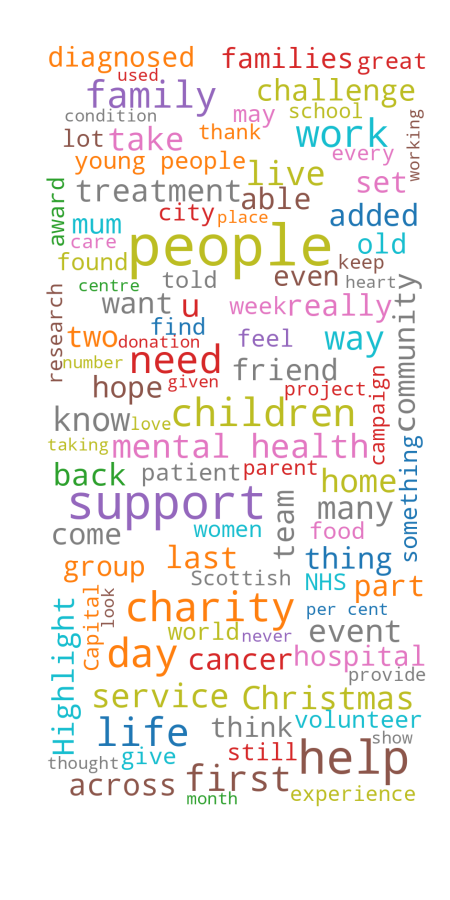}
    \vspace{-2.25\baselineskip}
    \caption{Cluster 32: ``Charity''}
    \end{subfigure}
    \begin{subfigure}[b]{.19\textwidth}
    \centering
    \captionsetup{justification=centering}
    \includegraphics[width=\textwidth]{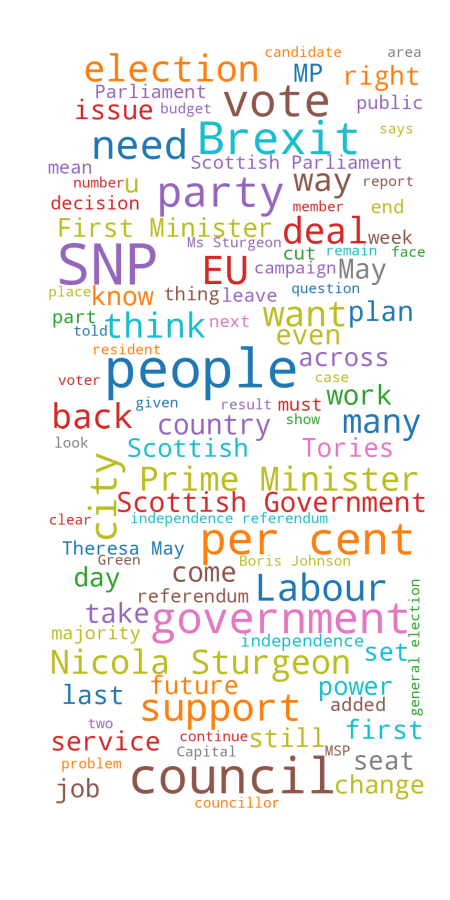}
    \vspace{-2.25\baselineskip}
    \caption{Cluster 33: ``Politics''}
    \end{subfigure}
    \begin{subfigure}[b]{.19\textwidth}
    \centering
    \captionsetup{justification=centering}
    \includegraphics[width=\textwidth]{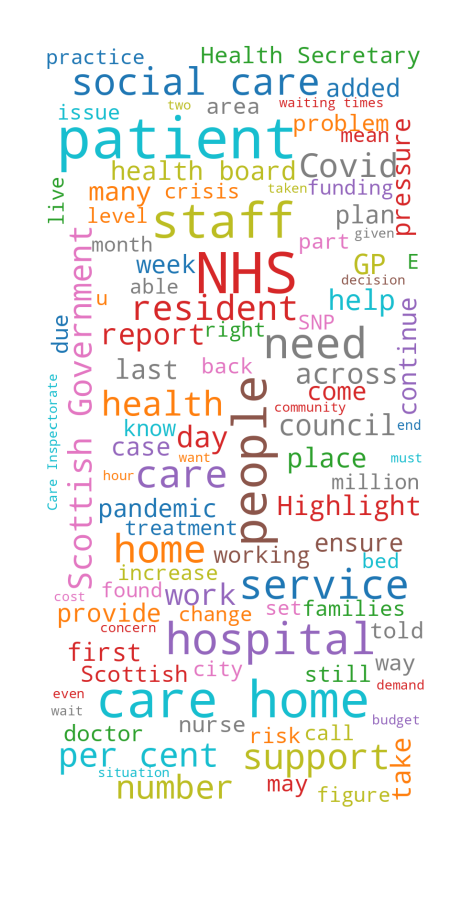}
    \vspace{-2.25\baselineskip}
    \caption{Cluster 34: ``Healthcare''}
    \end{subfigure}
    \begin{subfigure}[b]{.19\textwidth}
    \centering
    \captionsetup{justification=centering}
    \includegraphics[width=\textwidth]{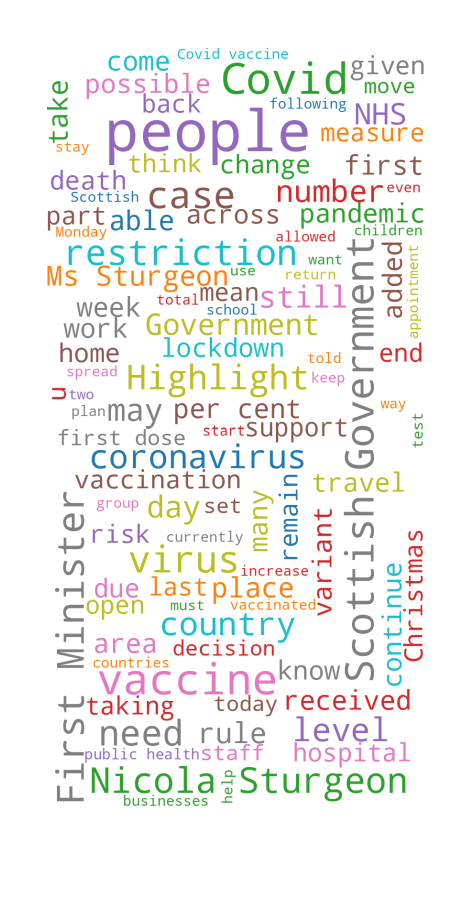}
    \vspace{-2.25\baselineskip}
    \caption{Cluster 35: ``Covid-19''}
    \end{subfigure}
    \begin{subfigure}[b]{.19\textwidth}
    \centering
    \captionsetup{justification=centering}
    \includegraphics[width=\textwidth]{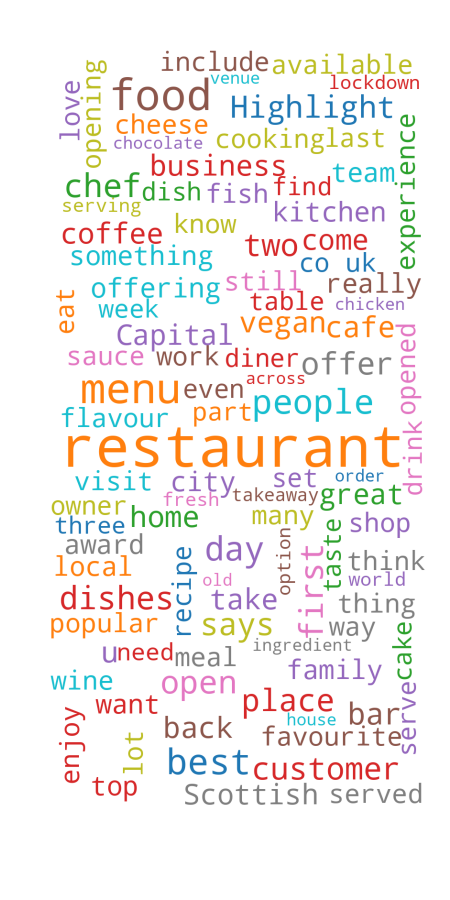}
    \vspace{-2.25\baselineskip}
    \caption{Cluster 36: ``Restaurants''}
    \end{subfigure}
    \begin{subfigure}[b]{.19\textwidth}
    \centering
    \captionsetup{justification=centering}
    \includegraphics[width=\textwidth]{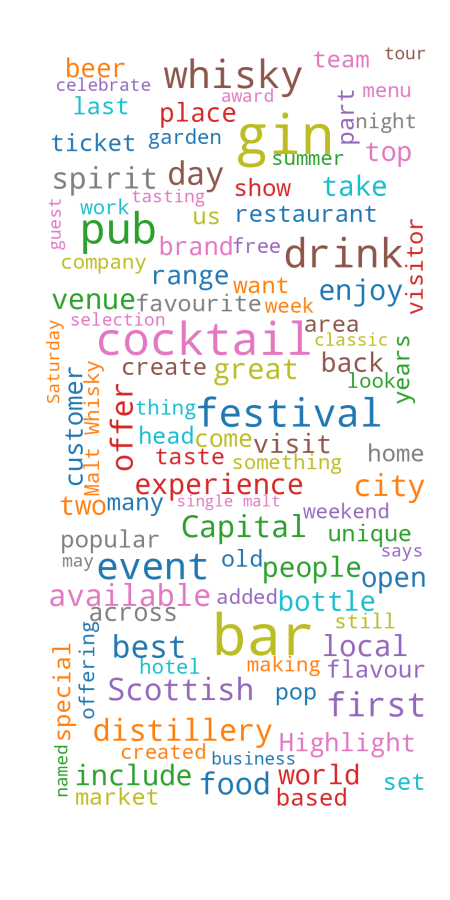}
    \vspace{-2.25\baselineskip}
    \caption{Cluster 37: ``Bars and pubs''}
    \end{subfigure}
    \begin{subfigure}[b]{.19\textwidth}
    \centering
    \captionsetup{justification=centering}
    \includegraphics[width=\textwidth]{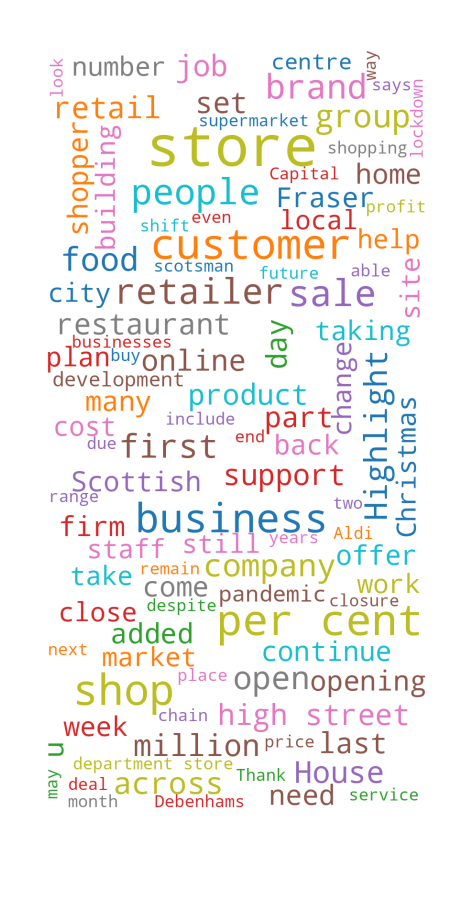}
    \vspace{-2.25\baselineskip}
    \caption{Cluster 38: ``Business and retail''}
    \end{subfigure}
    \begin{subfigure}[b]{.19\textwidth}
    \centering
    \captionsetup{justification=centering}
    \includegraphics[width=\textwidth]{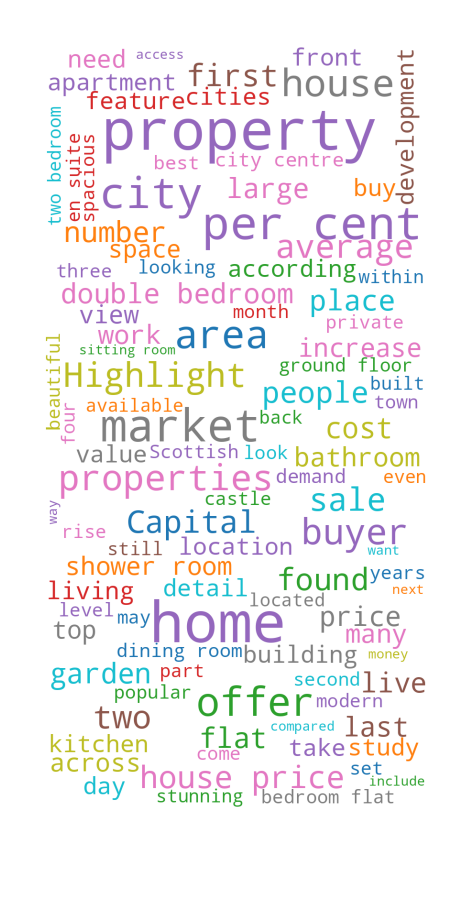}
    \vspace{-2.25\baselineskip}
    \caption{Cluster 39: ``Property market''}
    \end{subfigure}
    \begin{subfigure}[b]{.19\textwidth}
    \centering
    \captionsetup{justification=centering}
    \includegraphics[width=\textwidth]{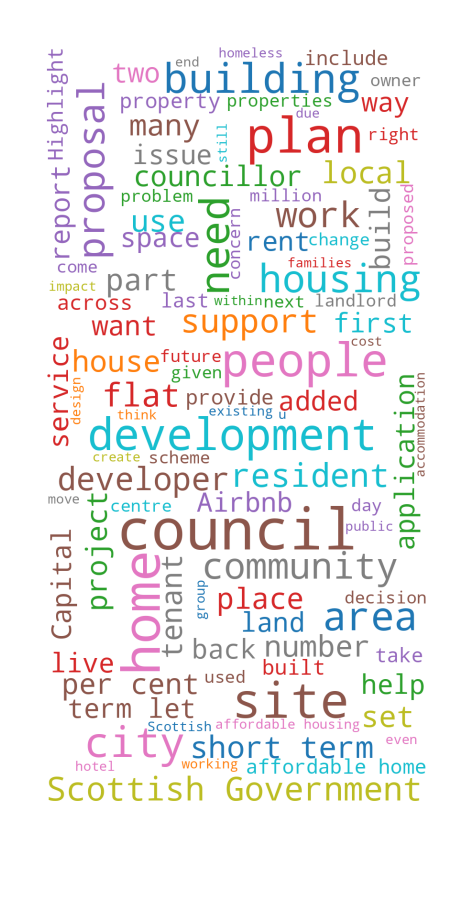}
    \vspace{-2.25\baselineskip}
    \caption{Cluster 40: ``City planning and development''}
    \end{subfigure}
    \begin{subfigure}[b]{.19\textwidth}
    \centering
    \captionsetup{justification=centering}
    \includegraphics[width=\textwidth]{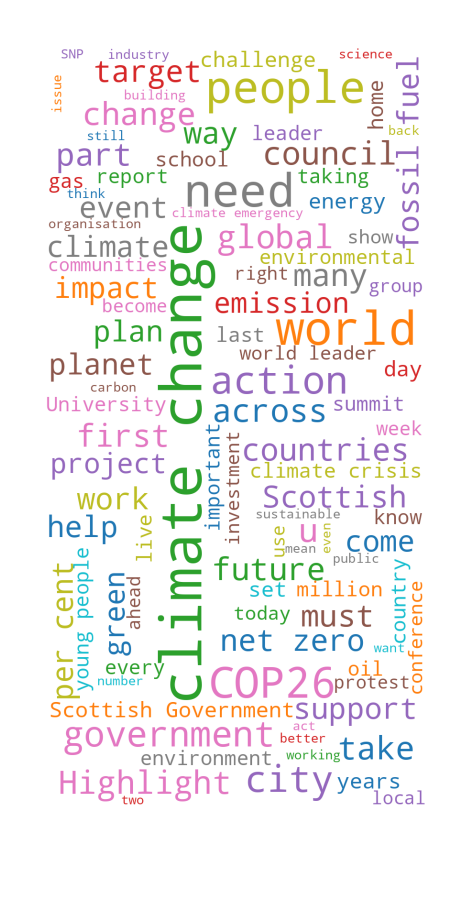}
    \vspace{-2.25\baselineskip}
    \caption{Cluster 41: ``Climate change''}
    \end{subfigure}
    \begin{subfigure}[b]{.19\textwidth}
    \centering
    \captionsetup{justification=centering}
    \includegraphics[width=\textwidth]{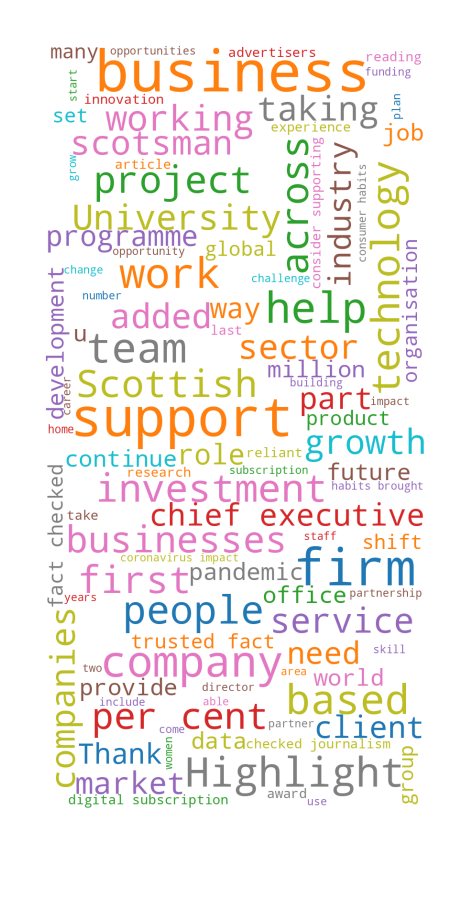}
    \vspace{-2.25\baselineskip}
    \caption{Cluster 42: ``Business''}
    \end{subfigure}
    \begin{subfigure}[b]{.19\textwidth}
    \centering
    \captionsetup{justification=centering}
    \includegraphics[width=\textwidth]{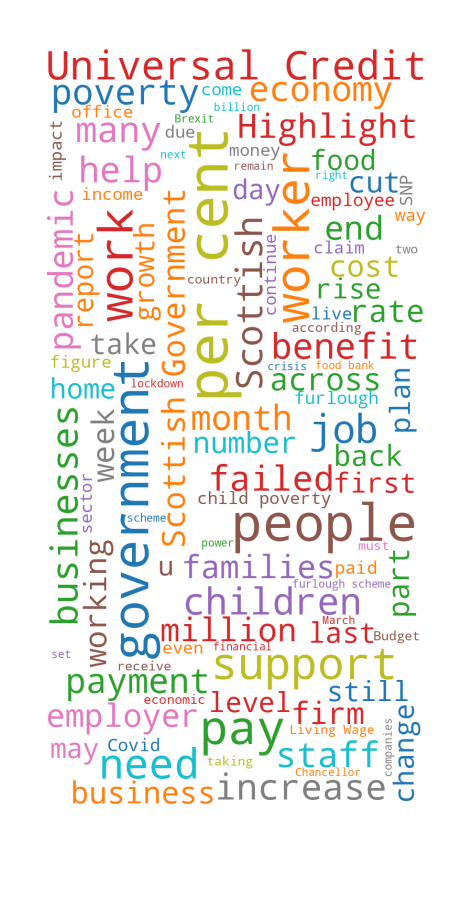}
    \vspace{-2.25\baselineskip}
    \caption{Cluster 43: ``Benefits''}
    \end{subfigure}
    \begin{subfigure}[b]{.19\textwidth}
    \centering
    \captionsetup{justification=centering}
    \includegraphics[width=\textwidth]{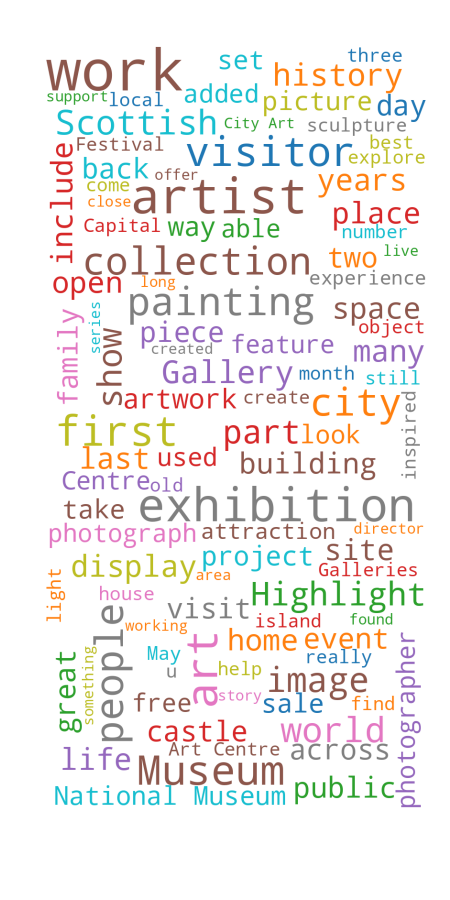}
    \vspace{-2.25\baselineskip}
    \caption{Cluster 44: ``Exhibitions''}
    \end{subfigure}
    \label{fig:clusters-3}
\end{figure}

\begin{figure}[t]
    \centering
    \begin{subfigure}[b]{.19\textwidth}
    \centering
    \captionsetup{justification=centering}
    \includegraphics[width=\textwidth]{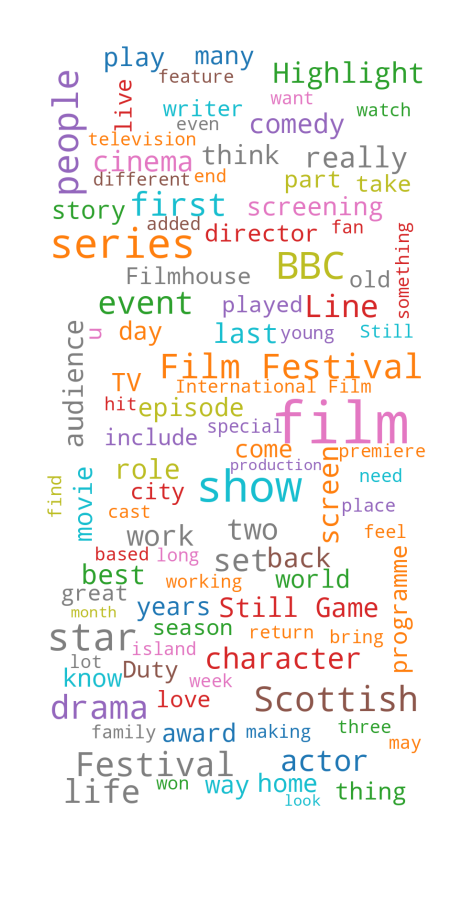}
    \vspace{-2.25\baselineskip}
    \caption{Cluster 45: ``Film''}
    \end{subfigure}
    \begin{subfigure}[b]{.19\textwidth}
    \centering
    \captionsetup{justification=centering}
    \includegraphics[width=\textwidth]{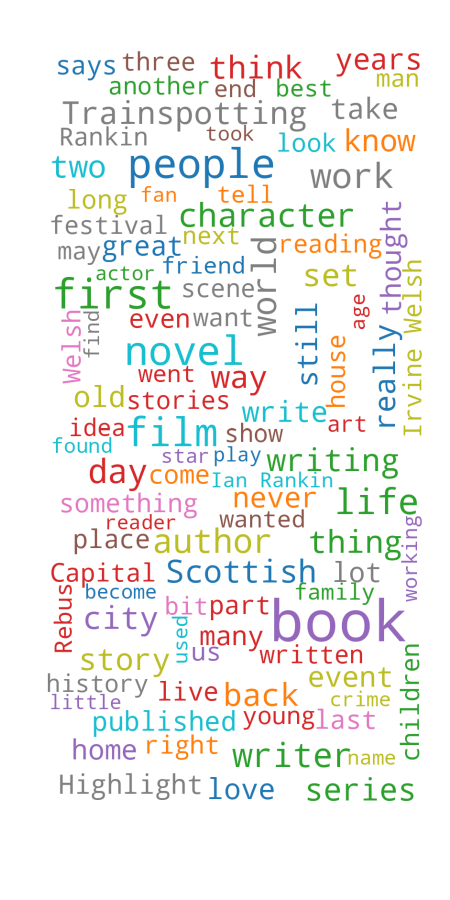}
    \vspace{-2.25\baselineskip}
    \caption{Cluster 46: ``Books''}
    \end{subfigure}
    \begin{subfigure}[b]{.19\textwidth}
    \centering
    \captionsetup{justification=centering}
    \includegraphics[width=\textwidth]{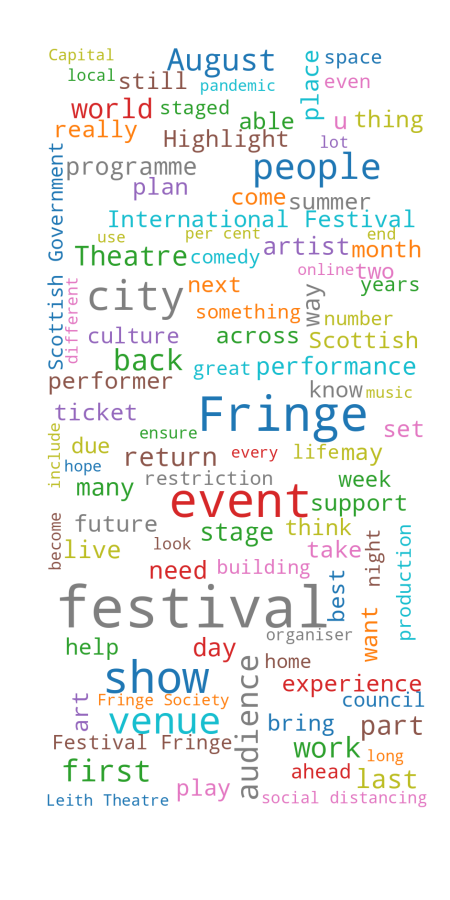}
    \vspace{-2.25\baselineskip}
    \caption{Cluster 47: ``Festival / Fringe''}
    \end{subfigure}
    \begin{subfigure}[b]{.19\textwidth}
    \centering
    \captionsetup{justification=centering}
    \includegraphics[width=\textwidth]{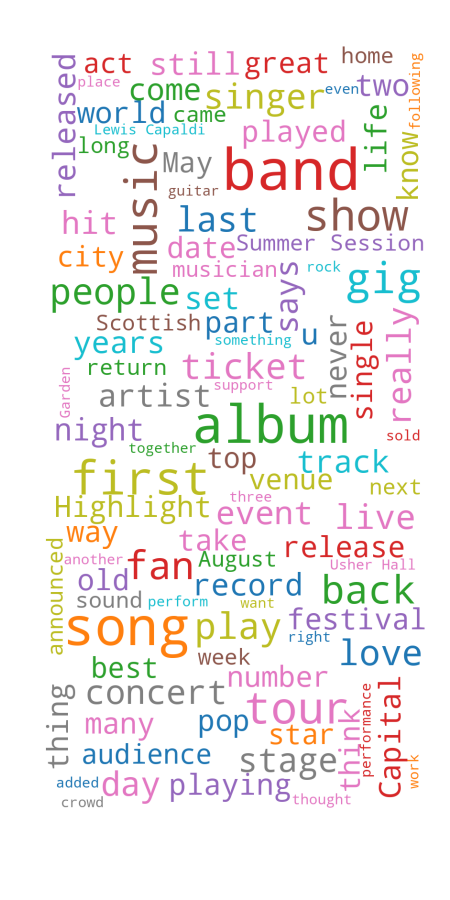}
    \vspace{-2.25\baselineskip}
    \caption{Cluster 48: ``Music''}
    \end{subfigure}
    \begin{subfigure}[b]{.19\textwidth}
    \centering
    \captionsetup{justification=centering}
    \includegraphics[width=\textwidth]{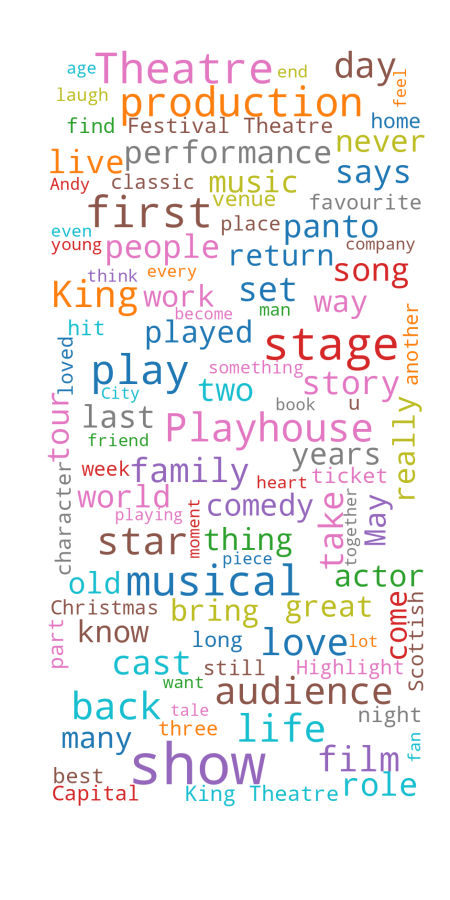}
    \vspace{-2.25\baselineskip}
    \caption{Cluster 49: ``Theatre''}
    \end{subfigure}
    \label{fig:clusters-4}
\end{figure}
\FloatBarrier
\section{Clustering Hierarchy}
\begin{figure}[h]
    \centering
    \includegraphics[width=.9\textwidth]{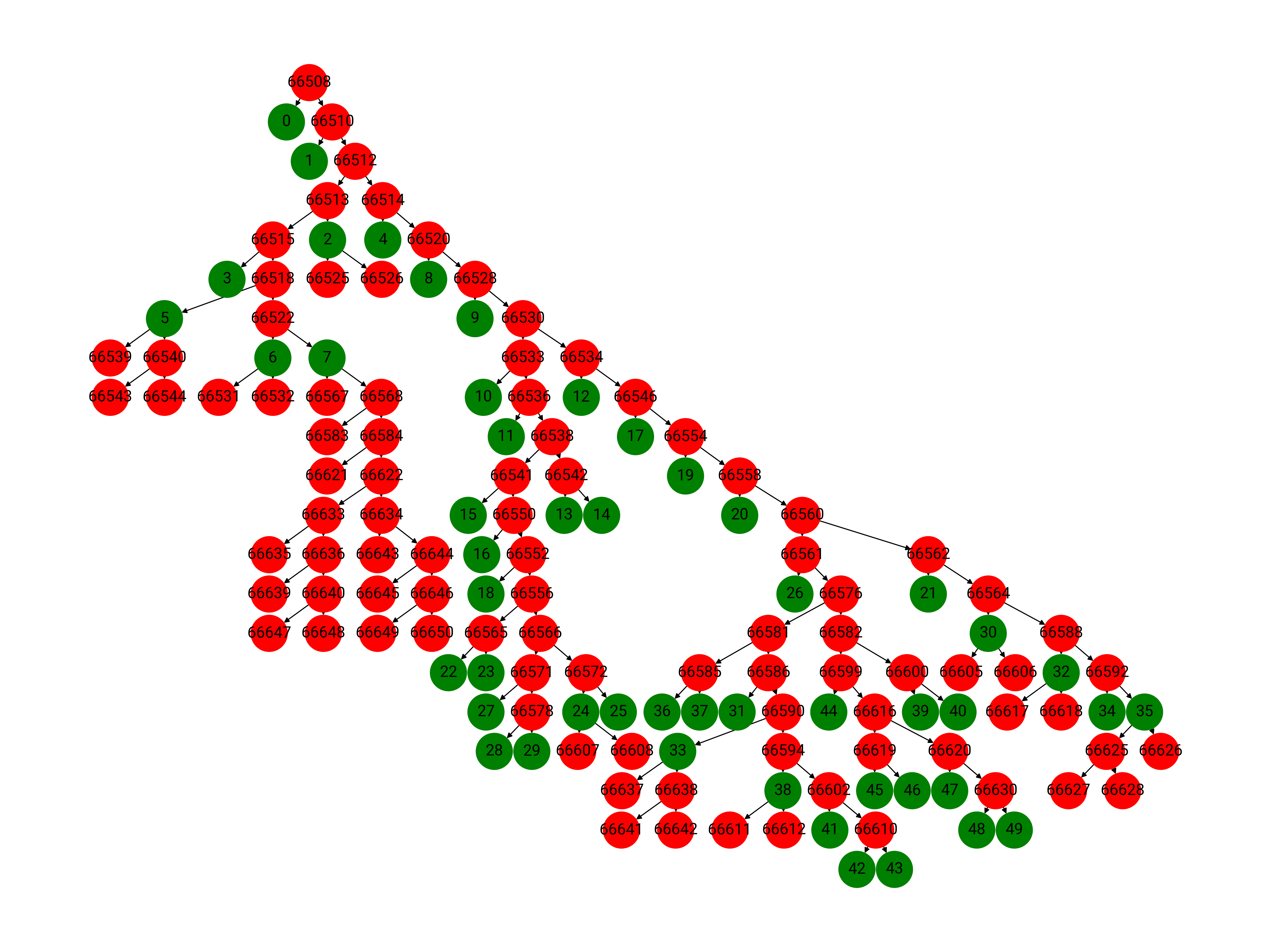}
    \caption{Hierarchy of clusters}
    \label{fig:hierarchy}
\end{figure}
\FloatBarrier
\section{Cross-validation Results}
\label{app:cross}
\subsection{Hyperparameter Selection}
\begin{table}
\centering
\resizebox{.9\textwidth}{!}{
\begin{minipage}{\textwidth}
\begin{tabular}{ccccc}
Macro F1 & \# Clusters & \# UMAP d & \# UMAP Nbrs & TF-IDF Vocab \\
\hline
78.09 & 51 & 10 &  5 & 20000  \\
77.97 & 58 & 20 &  5 & 10000  \\
77.74 & 51 & 10 &  5 & 10000  \\
77.18 & 68 &  2 &  5 & 10000  \\
77.10 & 51 & 20 &  5 & 20000  \\
76.89 & 63 &  2 &  5 & 20000  \\
76.62 & 59 &  2 & 15 & 10000  \\
76.36 & 44 & 20 & 15 &  5000  \\
76.13 & 64 &  2 & 15 & 20000  \\
76.13 & 50 & 20 &  5 &  5000  \\
74.48 & 65 &  2 & 30 & 10000  \\
74.28 & 55 &  2 &  5 &  5000  \\
73.95 & 65 &  2 & 30 & 20000  \\
73.64 & 44 & 20 & 15 & 20000  \\
73.15 & 51 & 10 &  5 &  5000  \\
72.92 & 64 &  2 & 30 &  5000  \\
72.45 & 43 & 20 & 30 & 10000  \\
72.34 & 54 & 10 & 15 &  5000  \\
72.27 & 45 & 20 & 30 &  5000  \\
71.55 & 43 & 10 & 30 &  5000  \\
71.43 & 48 & 20 & 30 & 20000  \\
70.27 & 44 & 10 & 30 & 10000  \\
69.51 & 53 & 10 & 15 & 10000  \\
68.54 & 48 & 10 & 15 & 20000  \\
66.48 & 69 &  2 & 15 &  5000  \\
66.30 & 46 & 10 & 30 & 20000  \\
62.39 & 60 & 20 & 15 & 10000  \\
\end{tabular}
\end{minipage}
}
\caption{Clustering results as measured by MacroF1 on the validation set for different choices of hyperparameters.}
\label{tab:crossvalidation}
\end{table}
We conduct a grid search over 27 parameter subsets.
 See \href{https://docs.google.com/spreadsheets/d/1kImpTcKQcOxe6bvmq6SDov_YCG78aqClEYmFVwVuQRk/edit#gid=614326044}{Spreadsheet for the data}.
We note that the Macro-F1 score varies a lot depending on the choice of hyperparameters with a range of values in $(62\%-78\%)$

\FloatBarrier

\section{Article Pair Annotation}
\label{app:annotation}

\begin{figure}[h]
    \centering
    \includegraphics[width=.65\columnwidth]{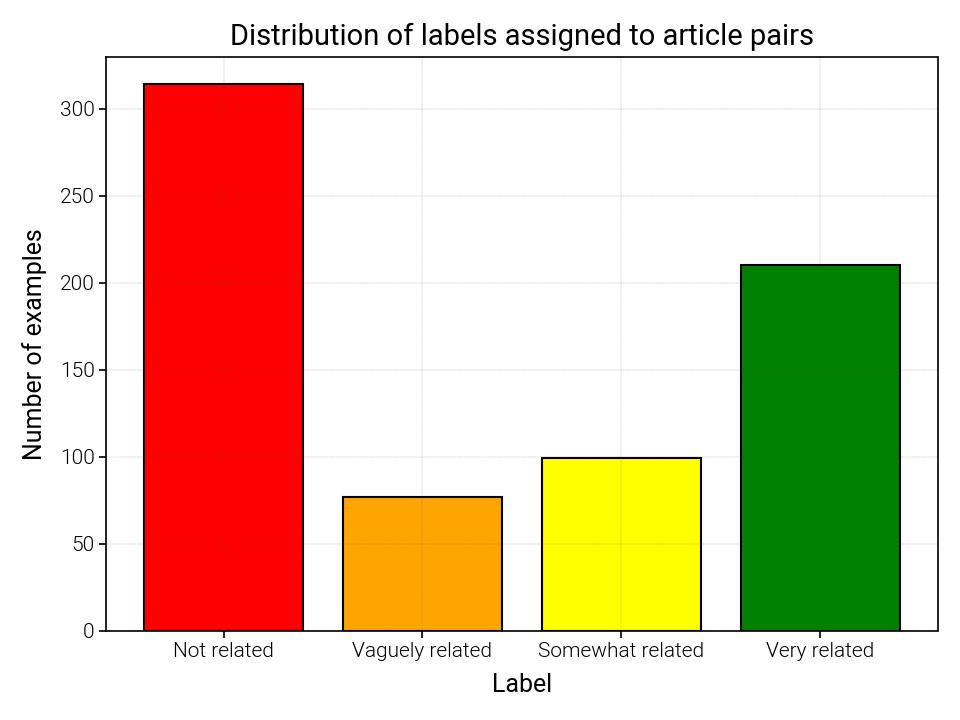}
    \caption{Article pair annotation: The number of article pairs falling within each level.}
    \label{fig:enter-label}
\end{figure}

Two articles talking about the same football team, e.g. Hibernian, are ``Very related''. An article on music and another on theatre are ``Somewhat related''. We used ``Vaguely related'' when a connection can be drawn but it is tenuous at best.

\FloatBarrier

\section{Article Pair Examples}
\begin{table}
\centering
\tiny
\resizebox{.9\textwidth}{!}{
\begin{minipage}{\textwidth}
\begin{tabular}{p{6cm}p{6cm}c}
\toprule
 Title of Article A & Title of Article B &  Score \\
\midrule
The three film studios at the heart of the Scottish screen industry... & The Scots actors who embraced change to bounce back from the pandem... &      3 \\
Five arts venues share £3m funding rescue package from Scottish Gov... & National Galleries plead for help to finish troubled project after ... &      3 \\
XXXXXXXXX roadworks: Four-way temporary traffic lights between XXXX... &                              XXXXX XXXX XXXXXX closed due to roadworks &      3 \\
Edinburgh Zoo giraffes: Pictures show the Capital's newest resident... &  Edinburgh Zoo welcomes three new endangered horses native to XXXXXXXX &      3 \\
Edinburgh Zoo bosses still planning for the return of gentle giant ... &      Edinburgh Zoo to pay half-price deal to keep pandas for two years &      3 \\
Scottish Canal Championships: New date for angling competition on X... &    Fishing: Alex Stewart from XXXXXXXXXX named in XXXXXXXX youth squad &      3 \\
XXXXXXXXX's Hogmanay film is seen by three million people in more t... & Basement Jaxx to headline XXXXXXXXX's Hogmanay festival with Ross B... &      3 \\
XXXXXXXX singer Roddy Woomble announces new solo album recorded dur... & Historic Colinton and Currie Pipe Band disbanded after 134 years pl... &      3 \\
Bilbo Baggins: Guitarist with XXXXXXXXX band who enjoyed 70s chart ... & Charlie Watts dies at 80: Trainspotting author Irvine Welsh leads c... &      3 \\
\midrule
XXXXXXXXX meal delivery business aimed at over-60s looks to become ... & Plans for XXXXX \& XXXX brewery at XXXXXX Watt University research p... &      2 \\
  ITV Viewpoint final episode pulled following Noel Clarke allegations & Daniel Radcliffe says Harry Potter author JK Rowling is 'an immense... &      2 \\
Leading Scottish theatres accused of shunning Scottish actors in ne... & Fantastic Beasts 3: JK Rowling's latest Harry Potter spin-off film ... &      2 \\
German-born composer of 'Highland Cathedral' anthem pens sequel 40 ... & XXXXXXX's Got Talent dancer who touched 'the lives of many' dies ag... &      2 \\
\midrule
Shannon Singh: Love Island star from XXXX opens up about time in th... & XXXXXXX's historic Kelvin Hall to become new base for TV entertainm... &      1 \\
Scottish event organisers and freelancers to get another £17.5m in ... & New film studio is finally set to help XXXXXXXXX raise its screen g... &      1 \\
Spectacular Snow Moon captured by Evening News reader atop city lan... & Foodies Festival: Tasty weekend festival set to return to XXXXXXXXX... &      1 \\
Crowds gather to pay tribute after death of teenage boxer Scott Martin & Edinburgh Leisure encourage locals to spring into summer with no jo... &      1 \\
\midrule
Royal Edinburgh Military Tattoo chief points finger of blame at XX ... & XXXXXXXXX social enterprise hair salon opens first of its kind hair... &      0 \\
XXXXXXXXXX XXXXXXXXX closed for weekend as search continues for mis... &              COP26 court plans in chaos as solitiors refuse to sign up &      0 \\
Forth One's Boogie in the Morning wins award at 'Oscars for XX radi... &                       XXXXXXXX's only porcupette born at Edinburgh Zoo &      0 \\
      Your chance to give your views on winter festivals- Kevin Buckle & And toddler made three for a ladies lunch with heart and soul - Hay... &      0 \\
XXXXX XXXXX angling proving popular - with some expert advice given... &         Hockey: Watsonians missing key men for Capital derby cup clash &      0 \\
XXXXXXXX's first social enterprise salon opens new hairdressers in ... & Rebooted festival has designs on seeing the city of XXXXXX in a new... &      0 \\
Appeal 9 years on: Family of former St Andrews University student w... & Two appear in court over £141,000 cannabis plant haul in XXXX XXXXX... &      0 \\
Scottish Parliament asked to right 'terrible miscarriage of justice... & XXXXXXXX launches bid to become 'world-class' hub for film, TV and ... &      0 \\
Pride Month: Virgin launch dedicated radio station for LGBT communi... & George IV Bridge Fire: Firefighter was hospitalised tackling XXXXXX... &      0 \\
Here's how we take this one-time opportunity to save the high stree... & We need the full story on late social worker Sean Bell - John McLellan &      0 \\
New film studio is finally set to help XXXXXXXXX raise its screen g... & Crunch talks on XXXXXXXXX Hogmanay to be held as organisers press a... &      0 \\
Edinburgh's Royal XXXXXXX XXXXXX transformed by 'invisible' art in ... & Scottish Ballet to make 'important' changes to The Nutcracker after... &      0 \\
Iconic ice-cream shop Luca's close its stores 'until further notice... & Watch: 'I'm not a cat,' says Texas lawyer during hilarious Zoom fil... &      0 \\
XXXXXXXXX's old Royal High School building: National Centre for Mus... & XXXXXX - The International Capital: Why I wrote my new book - Angus... &      0 \\
'Lake of pee' warning over XXXXX XXXXXXX restaurant's bid to expand... & Photographer spent six months magnifying seeds and fruits at the Ro... &      0 \\
Rebooted festival has designs on seeing the city of XXXXXX in a new... & Covid XXXXXXXX: XXXXXXXXX's Hogmanay revellers must provide proof o... &      0 \\
Jason Leitch suggests outdoor events will return first in XXXXXXXX ... & XXXXXXXXX's film studio saga is finally set to be turned into a scr... &      0 \\
Calcutta Cup: SRU says it 'fully supports' the fight against racism... & JK Rowling refuses to be intimidated after trans-rights activists r... &      0 \\
XXXXXXXXX's first female-centric bookstore proves to be a huge hit ... & XXXXXXXXX and the XXXXXXXX: Explore local and be a tourist in your ... &      0 \\
           Should drug use be decriminalised in XXXXXXXX? - your views & 'I've travelled the length of this country and XXXXXXXXX roads are ... &      0 \\
Porpoise washed up on beach following humpback sightings in XXXXX X... & Fire safety message pitched at campers in XXXXXXXX as number of gra... &      0 \\
    Sam Heughan says he would ' jump' at the chance to play James Bond & Scottish universities advised to postpone freshers' week when stude... &      0 \\
Edinburgh Council urged to fix security failings after arson attack... & XXXXXXXXX XXXX Council: Tory member John McLellan will not stand at... &      0 \\
\bottomrule
\end{tabular}
\end{minipage}
}
\caption{Analysis of annotations of Article Pairs which were classified as outliers.}
\label{tab:outliers}
\end{table}

\end{document}